\documentclass[12pt, draftclsnofoot, onecolumn]{IEEEtran}
\usepackage{graphicx}
\usepackage{epsfig}
\usepackage{algorithm}
\usepackage{algorithmic}
\usepackage{ifmtarg}
\usepackage{cite}
\usepackage{amssymb}
\usepackage{amsthm}
\usepackage[fleqn]{amsmath}
\usepackage{amsmath}
\usepackage{color}
\usepackage{bbm}
\usepackage{cite}
\usepackage{epstopdf}
\usepackage[table,xcdraw]{xcolor}
\usepackage{flushend}
\newtheorem{theorem}{\textbf{\text{Theorem}}}
\newtheorem{lemma}{\textbf{\text{Lemma}}}

\newtheorem{corollary}{Corollary}
\newtheorem{assumption}{Assumption}
\newtheorem{remark}{Remark}
\usepackage{flushend}
\usepackage{subcaption}

\begin{document}
\title{Flexible Design for $\alpha$-Duplex Communications in Multi-Tier Cellular Networks}
\author{
\IEEEauthorblockN{\large Ahmad AlAmmouri, Hesham ElSawy, and Mohamed-Slim Alouini}
\thanks{The authors are with Computer, Electrical, and Mathematical Sciences and Engineering (CEMSE) Divison, King Abdullah University of Science and Technology (KAUST), Thuwal, Makkah Province, Saudi Arabia. (Email: \{ahmad.alammouri, hesham.elsawy, slim.alouini\}@kaust.edu.sa)}
\thanks{Part of this work will be presented in IEEE International Conference on Communications (ICC'16) \cite{Harvesting2016AlAmmouri}.}}

\maketitle
\thispagestyle{empty}
\pagestyle{empty}

\begin{abstract}

Backward compatibility is an essential ingredient for the success of new technologies. In the context of in-band full-duplex (FD) communication, FD base stations (BSs) should support half-duplex (HD) users' equipment (UEs) without sacrificing the foreseen FD gains. This paper presents flexible and tractable modeling framework for multi-tier cellular networks with FD BSs and FD/HD UEs. The presented model is based on stochastic geometry and accounts for the intrinsic vulnerability of uplink transmissions. The results show that FD UEs are not necessarily required to harvest rate gains from FD BSs. In particular, the results show that adding FD UEs to FD BSs offers a maximum of $5\%$ rate gain over FD BSs and HD UEs case if multi-user diversity is exploited, which is a marginal gain compared to the burden required to implement FD transceivers at the UEs' side. To this end, we shed light on practical scenarios where HD UEs operation with FD BSs outperforms the operation when both the BSs and UEs are FD and we find a closed form expression for the critical value of the self-interference attenuation power required for the FD UEs to outperform HD UEs.

\end{abstract}


\section{Introduction}
\normalsize
Time division duplexing (TDD) and frequency division duplexing (FDD) are the commonly used techniques to protect receivers from their overwhelming self-interference (SI). This implies that the resources (i.e., time or frequency) are divided between forward and reverse links, which creates a performance trade-off between them. SI cancellation (SIC) eliminates such trade-off via in-band full-duplex (FD) communication, which gives the forward and reverse links the opportunity to simultaneously utilize the complete set of resources~\cite{Full2013Bharadia,Applications2014Hong,InBand2014Sabharwal,A2015Kim,On2014Alves}. 
FD transceivers are capable of sufficiently attenuating (up to -110 dB \cite{Full2015Goyal}) their own interference (i.e., SI) and simultaneously transmit and receive on the same channel, which offers higher bandwidth (BW) for FDD systems and longer transmission time for TDD systems. Consequently, FD communication improves the performance of both the forward and reverse links, in which the improvement depends on the efficiency of SIC.

Leveraging FD communication to large-scale networks, SI is not the only bottleneck due to the increased mutual interference when compared to the half-duplex (HD) case. This is because each FD link contains two active transmitters while each HD link contains one active transmitter and one passive receiver. Therefore, rigorous studies that capture the effect of the network interference on FD communication are required to draw legitimate conclusions about its operation in large-scale setup. In this context, stochastic geometry can be used to model FD operation in large scale networks and understand its behavior \cite{Stochastic2013ElSawy}. Stochastic geometry succeeded to provide a systematic mathematical framework for modeling both ad-hoc and cellular networks \cite{Stochastic2012Haenggi,A2011Andrews,Stochastic2013ElSawy,Stochastic2015Lu}.

Despite the higher interference injected into the network, recent studies have shown that FD communications outperform HD communications in large scale setup if sufficient SIC is achieved. For instance, the asymptotic study in  \cite{Does2015Xie} shows a maximum improvement of $80\%$ rate gain, which monotonically decreases in the link distance, for FD communication over the HD case. A more realistic ad-hoc network setup in \cite{Throughput2015Tong2} shows that FD offers an average of $33\%$ rate gain when compared to the HD operation. In the case of cellular networks, \cite{Hybrid2015Lee} shows around $30\%$ improvement in the total rate for FD when compared to the HD case. The authors in \cite{Throughput2016Goyal} show that the increase of aggregate interference in FD networks creates a trade-off between the average spectral efficiency and the coverage probability. However, \cite{Full2015Goyal} reveals that the FD gains in cellular networks are mainly confined to the DL due to the high disparity between uplink (UL) and downlink (DL) transmission powers. Furthermore, the authors in \cite{Limits2015Tsiky, AlAmmouri2015Inband,Interference2016Randrianantenaina,Can2016ElSawy} show that when a constrained power control is employed in the UL, the FD communication gains in the DL may come at the expense of high degradation in the UL. The authors in \cite{Limits2015Tsiky} advise to use FD communications in small cell tiers such that the users' equipment (UEs) and base stations (BSs) have comparable transmit powers. For FD operation in macro tiers with high disparity between UL and DL transmit powers, the authors in \cite{AlAmmouri2015Inband} advocate using pulse shaping along with partial overlap between UL and DL spectrum to neutralize DL to UL interference and avoid deteriorating UL rate. With pulse shaping and partial UL/DL overlap, \cite{AlAmmouri2015Inband} shows a simultaneous improvement of $33 \%$ and $28\%$ in the UL and DL, respectively. It ought to be mentioned that, in addition to the UL/DL transmit power disparity, the asymmetric UL/DL traffic that naturally exists in practical cellular networks imposes another challenge to the FD operation~\cite{Throughput2015Mahmood}. 

To harvest the aforementioned gains, FD transceivers are required on both sides of each link. However, cellular networks operators can only upgrade their BSs and do not have direct access to upgrade UEs. Furthermore, the high cost of FD transceivers, in terms of complexity, power consumption and price, may impedes their penetration to the UEs' domain. Therefore, techniques to achieve FD gains in cellular networks with FD BSs and HD UEs are required. In this context, 3-nodes topology (3NT) is proposed in \cite{Full2014Sundaresan, Hybrid2015Lima, Analyzing2013Goyal,Full2015Mohammadi,Outage2015Psomas,Throughput2016Goyal} to harvest FD gains by serving two HD UEs within each FD BS. In 3NT, the BSs have SIC capabilities and can simultaneously serve HD UL and HD DL users on the same channels. That is, each BS can merge each UL/DL channel pair into a larger channel and reuse that channel to serve an UL and a DL users simultaneously. The studies in \cite{Hybrid2015Lima,Full2014Sundaresan,Analyzing2013Goyal,Full2015Mohammadi} show the potential of 3NT to harvest HD gains. However, the results in \cite{Full2014Sundaresan} are based on simulations, and the results in \cite{Analyzing2013Goyal,Full2015Mohammadi,Outage2015Psomas,Hybrid2015Lima} are based on a simplistic system models.

In this paper, we present a unified mathematical framework, based on stochastic geometry, to model 3NT (i.e., FD BSs and HD users) and 2-nodes topology (2NT) (i.e., FD BSs and FD UEs) in multi-tier cellular networks. The proposed mathematical framework is then used to conduct rigorous comparison between 3NT and 2NT. Different from \cite{Analyzing2013Goyal,Full2015Mohammadi,Outage2015Psomas}, the presented system model accounts for the explicit performance of UL and DL for cell center users (CCUs) and cell edge users (CEUs) in a multi-tier cellular network. It also captures more realistic system parameters than \cite{Analyzing2013Goyal,Full2015Mohammadi,Outage2015Psomas} by accounting for pulse-shaping, matched filtering, UL power control, maximum power constraint for UEs, UEs scheduling, and the different BSs' characteristics in each network tier. When compared to \cite{AlAmmouri2015Inband}, the proposed framework considers a multi-tier network with different FD topologies (i.e., 2NT and 3NT),  flexible association, different path-loss exponents between different network elements, and incorporate uncertainties in the SIC.  However, we exploit the fine-grained duplexing strategy proposed in \cite{AlAmmouri2015Inband} that allows partial overlap between the UL and DL channels, which is denoted as $\alpha$-duplex ($\alpha$D) scheme. The parameter $\alpha \in [0,1]$ controls the amount of overlap between UL and DL channels and captures the HD (at $\alpha=0$) and FD (at $\alpha=1$) as special cases. Beside being used to optimize the spectrum allocation to the UL and DL, the parameter $\alpha$ shows the gradual effect of the interference induced via FD communication on the system performance, and to optimize the amount of the overlap between UL and DL channels. The results show that 3NT can achieve close performance (within 5$\%$) when compared to the 2NT with FD UEs that have efficient SIC if multi-user diversity and UEs scheduling are exploited. On the other hand, if the FD UEs in the 2NT have poor SIC, the 3NT achieves a better performance. In both cases, it is evident that network operators do not need to bear the burden of implementing SIC in the UEs to harvest FD gains.

The rest of the paper is organized as follows: in Section II, we present the system model and methodology of the analysis. In Section III, we analyze the performance of the $\alpha$-duplex system. Numerical and simulation results with discussion are presented in Section IV before presenting the conclusion in Section V.

\textit{\textbf{Notations}}: $\mathbb{E} [.]$ denotes the expectation over all the random variables (RVs) inside $[.]$, $\mathbb{E}_{x} [.]$ denotes the expectation with respect to (w.r.t.) the RV $x$, $\mathbbm{1}_{\{.\}}$ denotes the indicator function which takes the value $1$ if the statement $\{.\}$ is true and $0$ otherwise, $.*$ denotes the convolution operator and $S^{*}$ denotes the complex conjugate of $S$, $\mathcal{L}_{x} (.)$ denotes the Laplace transform (LT) of the probability density function (PDF) of RV $x$ and \textit{Italic} letters are used to distinguish variables from constants. 

\section{System Model}
\subsection{Network Model}\label{sec:NetModel}
A $K$-tier cellular network is considered, in which the BSs\footnote{In this work we assume that both the BSs and the UEs are equipped with a single antenna. Combining FD with multiple-input and multiple-output (MIMO) transmitters is covered in \cite{Full2015Atzeni,Directional2016Psomas}. } in each tier are modeled via an independent homogeneous 2-D Poisson point processes (PPPs) \cite{Stochastic2012Haenggi} $\Phi^{(k)}_{{\rm d}}$, where $k \in \{1,,2,...,K \}$, with intensity $\lambda_k$. The location of the $i^{th}$ BS in the $k^{th}$ tier is denoted by $x_{k,i} \in \mathbb{R}^2$. Beside simplifying the analysis, the PPP assumption for abstracting cellular BSs is verified by several experimental studies \cite{Stochastic2012Haenggi, A2011Andrews,Spatial2013Guo}. UEs are distributed according to a PPP  $\Phi_{\rm u}$, which is independent from the BSs locations, with intensity $\lambda_{\rm u}$, where $\lambda_{\rm u}\gg \sum\limits_{k=1}^{K}\lambda_k$.  Within each tier, all BSs transmit with a constant power $P_{\rm d}^{(k)}$, however, the value of $P_{\rm d}^{(k)}$ varies across different tiers.  In contrast, UEs employ a truncated channel inversion power control with maximum transmit power constraint of $P_{\rm{u}}$ \cite{On2014ElSawy}. That is, each UE compensates for its path-loss to maintain a tier-specific target average power level of $\rho^{(k)}$ at the serving BS. UEs that cannot maintain the threshold $\rho^{(k)}$ transmit with their maximum power $P_{\rm{u}}$.  UEs who can keep the threshold $\rho^{(k)}$, are denoted as cell center users (CCUs), while UEs who transmit with their maximum power are denoted as cell edge users (CEUs)\cite{Load2014AlAmmouri}.

The power of all transmitted signals experiences a power law path loss attenuation with exponent $\eta>2$. Due to the different relative antenna heights and propagation environments, we discriminate between the path loss exponent for the paths between two BSs (DL to UL interference), two UEs (UL to DL interference), and a BS and a UE (UL to UL interference), which are respectively denoted by $\eta_{\rm du},\eta_{\rm ud}$, and $\eta_{\rm uu}$, as shown in Fig. \ref{fig:Network1}. Assuming channel reciprocity, the path loss exponent between a BS and a UE (i.e., DL to DL interference), denoted by $\eta_{\rm dd}$, is equivalent to the one between a UE and a BS (i.e., UL to UL interference) $\eta_{\rm uu}$, and hence, both symbols are used interchangeably\footnote{We assume that the path-loss exponents in each direction is different but equivalent in all tiers. Assuming equal path-loss exponents in all tiers is a common simplifying assumption in the literature.\cite{On2014ElSawy,Modeling2012Dhillon,Uplink2009Chandrasekhar}.}. Also, Rayleigh fading channels are assumed such that the channels power gains are independent and identically distributed (i.i.d) exponential RVs with unit means\footnote{Extending the results to capture other fading models can be done following \cite{Average2013Renzo,Modeling2016AlAmmouri}.}.

\subsection{Operation Modes and Spectrum Allocation}
\begin{figure}[t]
\centerline{\includegraphics[width=  4.5in]{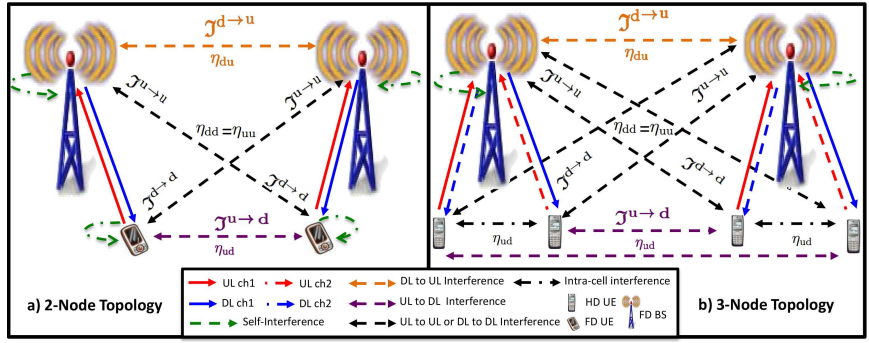}}
\caption{\, Channel allocation, interference types, and path-loss exponents for a) 2NT and b) 3NT.}
\label{fig:Network1}
\end{figure}

\begin{figure*}[t!]
    \centering
    \begin{subfigure}[t]{0.4\textwidth}
        \centerline{\includegraphics[width= 2.2 in]{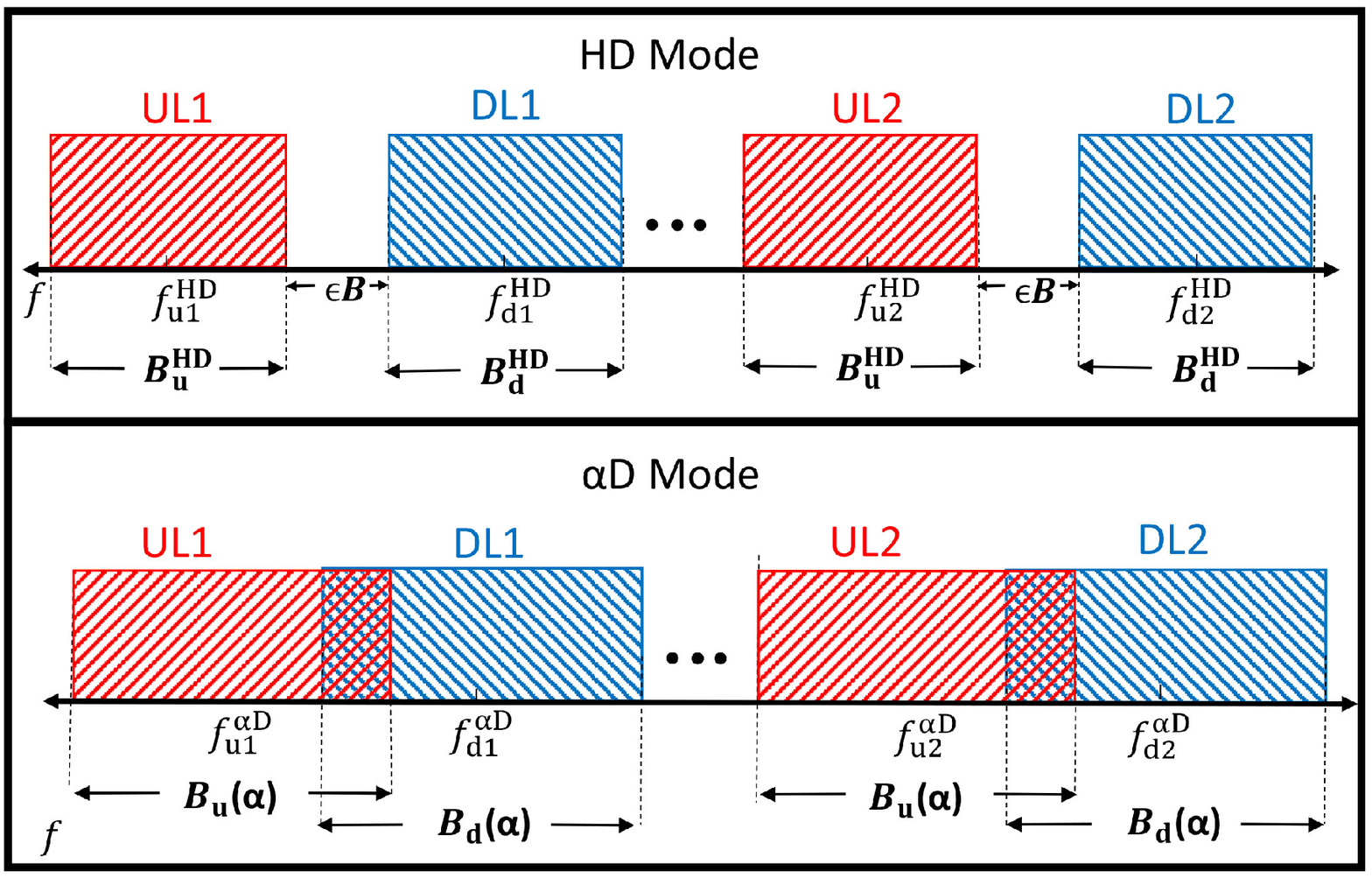}}\caption{\, Frequency bands allocation.}
\label{fig:Network2}
    \end{subfigure}%
    ~ 
    \begin{subfigure}[t]{0.4\textwidth}
       \centerline{\includegraphics[width=  1.6 in]{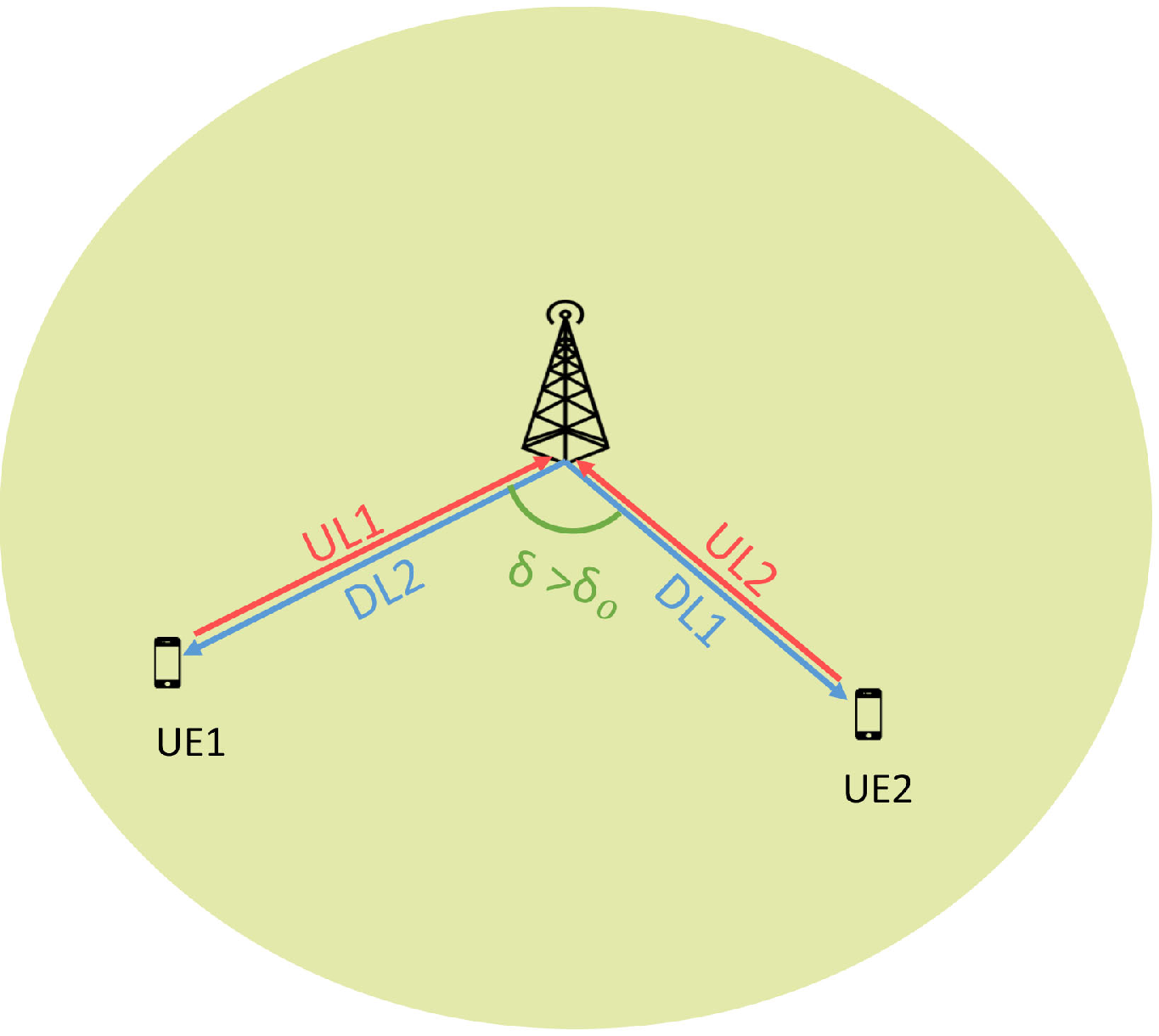}}\caption{\, UEs' scheduling in 3NT.}
\label{fig:Network4}
    \end{subfigure}
    \caption{Frequncy Bands allocation and UEs' scheduling.}\label{fig:association}
\end{figure*}

We consider a fine grained $\alpha$D scheme that allows partial overlap between UL and DL channels and captures the FD and HD as special cases. We denote the BWs used in the HD case in the UL and DL, respectively, as $B^{\rm HD}_{\rm{u}}$ and $B^{\rm HD}_{\rm{d}}$, in which $B^{\rm HD}_{\rm{u}}$ and $B^{\rm HD}_{\rm{d}}$ are not necessarily equal. To avoid adjacent channel interference, the BSs utilize a guard band of $\epsilon B$  between each UL-DL pair of bands, where $B= {{\rm min} (B^{\rm HD}_{\rm d},B^{\rm HD}_{\rm u})}$\footnote{ The scheme proposed in \cite{AlAmmouri2015Inband} is captured by setting $\epsilon$ to zero, since no guard bands are assumed there.}. As shown in Fig. \ref{fig:Network2}, the BW used in the $\alpha$D DL is $B_{\rm d}(\alpha) = B^{\rm HD}_{\rm{d}}+ \alpha (\epsilon +1) B$, and in the $\alpha$D UL is $B_{\rm u}(\alpha)=B^{\rm HD}_{\rm{u}}+ \alpha (\epsilon +1) B$. Note that the parameter $\alpha$ controls the partial overlap between the UL and DL frequency bands. Also, the HD and FD modes are captured as special cases by setting $\alpha$ to 0 and 1, respectively. It is assumed that each tier has its own duplexing parameter $\alpha_k$, which is used by all BSs within that tier.

\normalsize

Without loss of generality, we assume that each BS has only two pairs of UL-DL channels that are universally reused across the network. For simplicity,  we assume that the two channel pairs are sufficiently separated in the frequency domain (i.e., $f^{\rm HD}_{\rm u1} < f^{\rm HD}_{\rm d1} \ll f^{\rm HD}_{\rm u2} <f^{\rm HD}_{\rm d2}$) to avoid adjacent channel interference between different UL-DL pairs. It is worth noting that the idealized rectangular frequency domain pulse shapes shown in Fig. \ref{fig:Network2} are used for illustration only. However, as discussed later, we use time-limited pulse shapes that impose adjacent channel interference due to the out of band ripples in the frequency domain. 

In the 2NT network, UEs have FD transceivers and can use the UL and DL belonging to the same UL-DL pair for their $\alpha$D operation. In contrast, 3NT UEs have HD transceivers and cannot transmit and receive on overlapping channels. Hence, each HD user is assigned his UL and DL channels from two different UL-DL pairs as shown in Fig. \ref{fig:Network1} and Fig. \ref{fig:Network2}. Consequently, 3NT UEs can benefit from the larger BW channels without SI. Note that the FD BSs in all cases as well as the FD UEs in the 2NT would experience SI as shown in Fig. \ref{fig:Network1}. In contrast, 3NT experience intra-cell interference on the DL direction due to the partial overlap between the UL channel of the one UE and the DL channel of the other UE.

To this end, we assume that the BSs exploit multi-user diversity to control intra-cell interference in 3NT by imposing a minimum separation angle constraint between users scheduled on the same channel as shown in Fig. \ref{fig:Network4}\footnote{More advanced and sophisticated scheduling and multi-user diversity techniques are postponed for future work.}. In sectored BSs, the value of $\delta_o$ can be estimated to a certain accuracy depending on the number of sectors. If the BSs' cannot estimate the angeles between users, then $\delta_o$ is set to zero and we refer to this case as random scheduling.

For the FD BSs and 2NT UEs, we denote the SI attenuation power as $\beta_{\rm u} h_{\rm s}$ and $\beta_{\rm d} h_{\rm s}$, respectively, where $\beta_{\rm u}$, $\beta_{\rm d}$ are positive constants representing the mean SIC power values in the UL and DL, respectively, and $h_{\rm s}$ follows a general unit mean distribution with PDF given by $f_{H_{\rm s}}(\cdot)$ which represents the uncertainty in SIC. Three special cases of interest for $f_{H_{\rm s}}(\cdot)$ are considered, namely, constant attenuation where  $f_{H_{\rm s}}(\cdot)$ is a degenerate distribution as in \cite{AlAmmouri2015Inband,Intra2015Yun,Hybrid2015Lee} and random attenuation where $f_{H_{\rm s}}(\cdot)$ is an exponential distribution as in \cite{Full2015Mohammadi} and Rician fading as in \cite{Full2015Atzeni} which captures the previous two cases as special cases. It is shown in \cite{Harvesting2016AlAmmouri} that all distributions leads to the same performance trends.

\subsection{UEs to BSs Association}\label{sec:Association}

\normalsize
\begin{figure*}[t!]
    \centering
    \begin{subfigure}[t]{0.3\textwidth}
        \centerline{\includegraphics[width= 1.8 in]{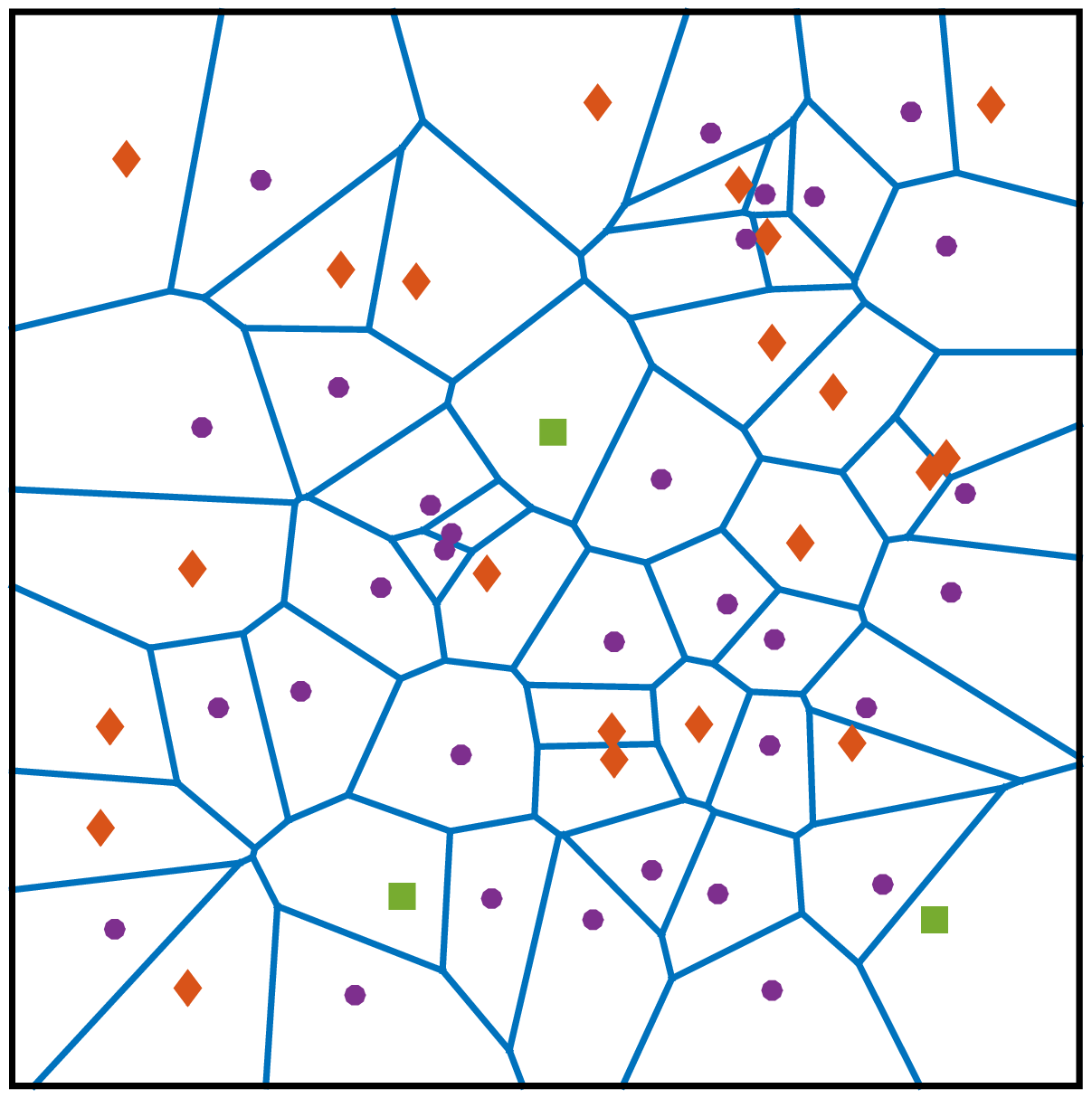}}\caption{\, $\tau_{j}=1$.}
\label{fig:CBA}
    \end{subfigure}%
    ~ 
    \begin{subfigure}[t]{0.3\textwidth}
       \centerline{\includegraphics[width=  1.8 in]{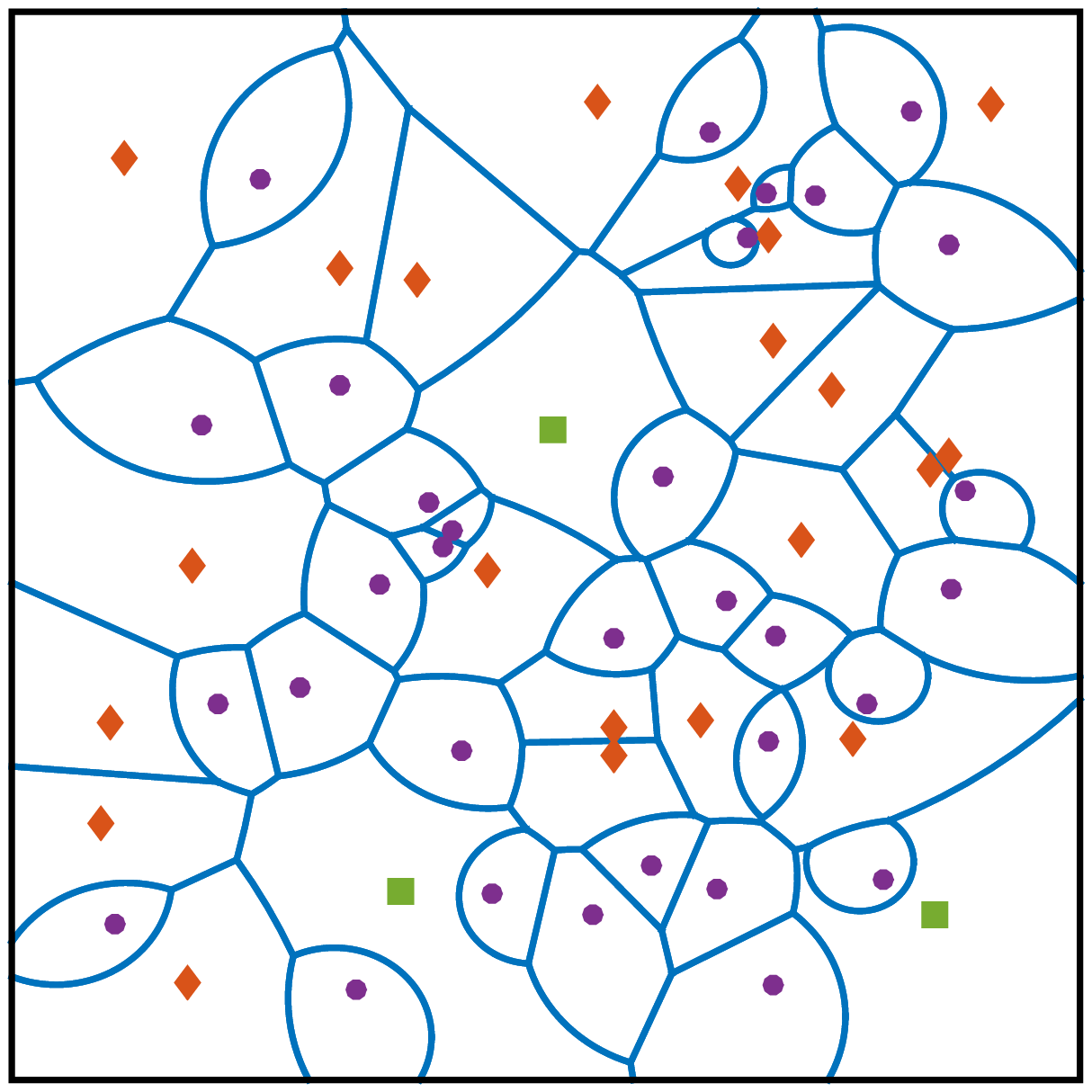}}\caption{\, $\tau_{j}=(P_{{\rm d}}^{(j)})^{\frac{-1}{\eta}}$.}
\label{fig:SBA}
    \end{subfigure}
    \caption{A realization of the associations areas assuming different association factors, where the green squares, diamonds, and circles represent macro, micro, and pico BSs, respectively.}\label{fig:association}\vspace{-0.8cm}
\end{figure*}

We consider a biased and coupled\footnote{Decoupled UL/DL association is analyzed using stochastic geometry in \cite{Joint2015Singh} for traditional HD multi-tier network, extending this analysis to decoupled association is postponed to future work.} BS-UE association scheme.  Biasing is used to enable flexible load balancing between tiers by encouraging UEs to connect to lower power BSs to balance the average load served by the tiers across the network~\cite{An2014Andrews,HetNets2013Elsawy}. We define a distance dependent biasing factor $\tau$ and assume that all BSs within the same tier have the same biasing factor. Hence, a UE connects to $k^{th}$ tier if $\{ \tau_{k} r_{k}<\tau_{i} r_{i} \ \forall \ i\in\{1,..,K\}, \  k\neq i \}$.

The used association scheme captures different association strategies as special cases. For example, if $\tau$ is set to the same value for all tiers, then closest BS association is considered, if $\tau_k=(P^{(k)}_{{\rm d}})^{\frac{-1}{\eta_{\rm dd}}}$, then the UE connects to the BS providing the highest received signal strength (RSS). Note that different association schemes changes the relative BSs' association areas across the tiers as shown in Fig. \ref{fig:association}, where three tiers network is shown with $10$W macro BSs, $5$W micro BSs, and $1$W, pico BSs\footnote{The values of the transmit powers are based on \cite{Hierarchical2011Jain}.}. In Fig. \ref{fig:CBA}, nearest BS association is considered, and hence, association areas are represented by Voronoi tessellation \cite{Generalized1986Geometriae}. In Fig. \ref{fig:SBA}, UE connects according to the RSS, in this case the association areas construct multiplicative weighted Voronoi tessellation (also denoted as circular tessellation) \cite{Generalized1986Geometriae}.

\subsection{Pulse Shaping}\label{sec:PulseShaping}

We employ time-limited pulse shapes\footnote{In this work, we focus on time-limited pulse shapes to avoid inter-symbol-interference (ISI), since including the effect of ISI will complicate the analysis much more. However, frequency-limited Nyquist pulses (e.g. root raised cosine) can be also used, since it protect the nodes from ISI. For more information on the effect of different pulse shapes on the $\alpha$-duplex scheme, refer to [17].}, denoted as $s(t,{\rm BW},b_v) \overset{\rm{FT}}{\longleftrightarrow} S(f,{\rm BW},b_v)$ with unit energy, where ${\rm BW}$ is the pulse null-to-null bandwidth, $b^{(k)}_{\rm d}$ and $b^{(k)}_{\rm u}$ indicate the pulse types used by the $k^{th}$ tier in the DL and UL, respectively. We assume a flexible pulse shaping scheme, where each tier has its own pulse shapes in the DL and UL, however, all BSs within the same tier use the same pulse shapes. To have a unified effective BW for all values of $\alpha_i$ in the $\alpha$D mode, the null-to-null BW of the pulse-shapes is kept equal to the channel BW. Hence, the pulse shapes are also functions of the parameter $\alpha_k$.

\subsection{Base-band Signal Representation}
For the sake of simple presentation, we use $\alpha_k$, $b^{(k)}_{\rm d}$ and $b^{(k)}_{\rm u}$ to denote the duplexing factor, the UL pulse shape, and the DL pulse shape, respectively in the $k^{th}$ tier. Also, we use $v,\bar{v},$ and $w$ to indicate the desired transmission, where $v,\bar{v},w\in \{{\rm d},{\rm u} \}, v\neq \bar{v}$, for DL and UL, respectively, and $i,k$ as BSs' tier index, where $i,k \in \{1,...,K\}$. Exploiting this notation, the received baseband signal at the input of the matched filter of a test transceiver in the $i^{th}$ tier (BS or UE) can be expressed~as

\small
\begin{align}
\!\!\!\!\!\!\!\!\!\!\!\! y_{v}^{(i)}(t)= & \Gamma_o \sqrt{P^{(i)}_{v_o} r_o^{-\eta_{vv}} h_o} s(t,B_{v}(\alpha_i),b^{(i)}_{v}) +\sum\limits_{k=1}^{K} \sum_{j \in \tilde{\Psi}^{(k)}_{\rm d}} \mathfrak{I}^{{\rm d}^{(k)} \rightarrow v^{(i)}}_{j}(t) + \sum\limits_{k=1}^{K} \sum_{j \in \tilde{\Psi}^{(k)}_{\rm u}} \mathfrak{I}^{{\rm u}^{(k)} \rightarrow v^{(i)}}_{j}(t)+ \mathfrak{I}_{{\rm s}_{v}}^{(i)}(t) + n(t).
\label{base_band_DL}
\end{align}\normalsize
\noindent{where}  $\Gamma_o$, $P^{(i)}_{{ v}_o}$, $r_o$, and $h_o$ denote  the intended symbol, transmit power, link distance, and channel power gains, respectively.  $\tilde{\Psi}^{(k)}_{\rm d} \subseteq  {\Psi}^{(k)}_{\rm d}$ is the set of interfering BSs in the $k^{th}$ tier, $ \mathfrak{I}^{{\rm d}^{(k)} \rightarrow v^{(i)}}_{j}(t)$  is the DL interference from the $j^{th}$ BS in $k^{th}$ tier, $\tilde{\Psi}^{(k)}_{\rm u} \subseteq  {\Psi}^{(k)}_{\rm u}$ is the set of interfering UEs in the $k^{th}$ tier, $ \mathfrak{I}^{{\rm u}^{(k)} \rightarrow v^{(i)}}_{j}(t)$  is the UL interference from $j^{th}$ UE connected to the $k^{th}$ tier,  $ \mathfrak{I}_{{\rm s}_{v}}^{(i)}(t)$ is the SI term affecting the $v$ direction, and $n(t)$ is a white complex Gaussian noise with zero mean and two-sided power spectral density $N_o/2$. The downlink and uplink interference are given by
\small
\begin{align} \label{inter1}
\mathfrak{I}^{{\rm d}^{(k)} \rightarrow v^{(i)}}_{j}(t)&=  \Gamma^{(k)}_{{\rm{d}}_{j}} s(t,B_{\rm d}(\alpha_i),b_{\rm d}^{(k)}) \sqrt{P^{(k)}_{\rm d} h^{(k)}_{{\rm d}_j} \left(r^{(k)}_{{\rm d}_j}\right)^{-\eta_{{\rm d}v}}} \exp \left( j 2 \pi \left(f^{(k)}_{\rm d}-f^{(i)}_{v}\right)t \right), \\
\mathfrak{I}^{{\rm u}^{(k)} \rightarrow v^{(i)}}_{j}(t)&=  \Gamma^{(k)}_{{\rm{u}}_{j}} s(t,B_{\rm u}(\alpha_i),b_{\rm u}^{(k)}) \sqrt{P^{(k)}_{{\rm u}_j} h^{(k)}_{{\rm u}_j} \left(r^{(k)}_{{\rm u}_j}\right)^{-\eta_{{\rm u}v}}}  \exp \left( j 2 \pi \left(f^{(k)}_{\rm u}-f^{(i)}_{v}\right)t \right).
 \label{inter2}
\end{align}\normalsize

\noindent{where} $\Gamma^{(k)}_{{\rm{d}}_{j}}$ and $\Gamma^{(k)}_{{\rm{u}}_{j}}$ denote the interfering symbol from the DL $j^{th}$ BS and interfering symbol from the UL  $j^{th}$ UE in the $k^{th}$ tier. Following the same interpretation of the subscripts and superscripts defined for the interfering symbols, $h^{(k)}_{{\rm d}_j}$ and $h^{(k)}_{{\rm u}_j}$ denote the DL and UL interfering channel gains, $P^{(k)}_{{\rm d}}$, and $P^{(k)}_{{\rm u}_j}$ denote the DL and UL interfering  transmit powers, $r^{(k)}_{{\rm d}_j}$, and $r^{(k)}_{{\rm u}_j}$ denote the DL and UL interfering link distances, and $f^{(k)}_{\rm d}$ and $f^{(k)}_{\rm u}$ denote the center frequencies of the DL and UL interfering frequency bands (see Fig. \ref{fig:Network1}). Note that the BS index is removed from the DL transmit power because all BSs in the same tier transmit with the same power. Similarly, the BS and UE indices are removed from the center frequencies  $f^{(k)}_{\rm d}$ and $f^{(k)}_{\rm u}$ because all elements in the same tier employ the same overlap parameter $\alpha^{(k)}$. The SI term in \eqref{base_band_DL} is given by

%$h_o$, $h^{(k)}_{{\rm d}_j}$ and $h^{(k)}_{{\rm u}_j}$ are the channel power gains between the test receiver and, respectively, the intended transmitter, interfering $j^{th}$ BS, and interfering $j^{th}$ UE in the $k^{th}$ tier, $P^{(i)}_{v_o}$ is the transmitted power by the desired transmitter, $r_o$ is the serving distance between the UE and its serving BS,  $\tilde{\Psi}^{(k)}_{\rm d} \subseteq   {\Psi}^{(k)}_{\rm d}$ is the set of interfering BSs in the $k^{th}$ tier, $\mathfrak{I}^{{\rm d}^{(k)} \rightarrow v^{(i)}}_{j}(t)$ is the DL interference from the $j^{th}$ BS in the $k^{th}$ tier, $\tilde{\Psi}^{(k)}_{\rm u} \subset   {\Psi}_{\rm u}$ is the set of interfering UEs associated with the $k^{th}$ tier, 

%The symbols transmitted by interfering network elements are abstracted via Gaussian codebooks as in \cite{A2009Shobowale}. 

% $h^{(k)}_{{\rm d}_j}$ and $h^{(k)}_{{\rm u}_j}$ are the channel power gains between the tagged receiver and the $j^{th}$ interfering BS and UE in the $k^{th}$ tier, $r^{(k)}_{{\rm d}_j}$ and $r^{(k)}_{{\rm u}_j}$ are the distances between the tagged receiver and the $j^{th}$ interfering BS and UE in the $k^{th}$ tier. $P^{(k)}_{{\rm u}_j}$ is the transmitted power of the
%$j^{\rm th}$ interfering UE associated with the $k^{th}$ tier, and $P^{(k)}_{\rm d}$ is the transmitted power of an interfering BS in the $k^{th}$ tier, 

\small
\begin{align}\label{eq:SIu1}
& \mathfrak{I}^{(i)}_{{\rm s}_{\rm u}}(t)= \Gamma_{s} \sqrt{\beta^{(i)}_{\rm u} h_s P_{\rm d}}s(t,B_{\rm d}(\alpha_i),b^{(i)}_{\rm d}) \exp \left( j 2 \pi \Delta f^{(i)} t \right).
\end{align}
\begin{align}\label{eq:SId1}
&\mathfrak{I}^{(i)}_{{\rm s}_{\rm d}}(t)= \left\{
	\begin{array}{ll}
		\Gamma_{s} \sqrt{\beta_{\rm d} h_s P_{{\rm u}_{o}}}s(t, B_{\rm u}(\alpha_i),b^{(i)}_{\rm u}) \exp \left(- j 2 \pi \Delta f^{(i)} t \right).     & {\rm 2NT}  \\
		0. & {\rm 3NT}
	\end{array}
\right.
\end{align}\normalsize
where, $\beta_{\rm d}$ represents the average attenuation power of the SI in the DL, $\beta^{(i)}_{\rm u}$ is the average attenuation power of the SI affecting a BS in the $i^{th}$ tier in the UL, hence, each tier can have a different SIC capability depending on the BSs' sizes and receivers complexity. $ P_{{\rm u}_{o}}$ is the transmit power of the tagged UE and

\begin{align}
\Delta f^{(i)}= f^{(i)}_{\rm u}-f^{(i)}_{\rm d},
\end{align}
which represents the difference between the UL and DL center frequencies in the $i^{th}$ tier. Note that this difference also depends on the chosen tier, since each tier can have a different duplexing factor $\alpha_i$ which leads to different UL/DL BWs and center frequencies.

\subsection{Methodology of Analysis} \label{method}

The analysis is conducted on a test transceiver, which is a BS for the UL and a UE for the DL, located at the origin and operating on a test channel pair. According to Slivnyak's theorem \cite{Stochastic2012Haenggi}, there is no loss of generality in this assumption. Also, there is no loss of generality to focus on a test channel pair as interferences on different bands are statistically equivalent. We asses the impact of FD communication via the outage probability and the  transmission rate. The outage probability is defined as
\begin{align}
\mathcal{O}(\theta)=\mathbb{P}\left\{{\rm SINR} <\theta\right\}.
\label{eq:Outage}
\end{align}

For the transmission rate \cite{Throughput2014Li}, we assume that the nodes transmit with a fixed rate regardless of the state of the channel (${{\rm BW}} \log_2 \left(1+\theta \right)$), hence, the transmission rate is defined as
\begin{align}\label{eq:Thr_Gen}
\mathcal{R}={{\rm BW}} \log_2 \left(1+\theta \right) \mathbb{P}\left\{{\rm SINR} \geq \theta\right\}.
\end{align}

In \eqref{eq:Thr_Gen}, the degraded SINR  is compensated by the increased linear BW term. Hence, \eqref{eq:Thr_Gen} can be used to fairly assess the performance of FD operation.  As shown \eqref{eq:Outage} and \eqref{eq:Thr_Gen}, both the outage probability and transmission rate are independent from the symbol structure and only depend on the SINR. Consequently, all transmitted symbols $\Gamma_o$,  $\Gamma^{(k)}_{{\rm{d}}_{j}}$ and $\Gamma^{(k)}_{{\rm{u}}_{j}}$ for all $\{k,j\}$ are abstracted to independent zero-mean unit-variance complex Gaussian random variables. Abstracting the symbols via Gaussian random variables have negligible effect on the signal-to-interference-plus-noise-ratio (SINR) distribution as shown in \cite{The2015Afify, Influence2005Giorgetti}. 

In the analysis, we start by modeling the effect of the matched and low-pass filtering on the baseband signal. Then, based on the base-band signal format after filtering, the expressions for the SINR in different cases (i.e., CCU-UL, CCU-DL, CEU-UL, and CEU-DL for 3NT, and 2NT) are obtained. The performance metrics in  \eqref{eq:Outage} and  \eqref{eq:Thr_Gen} are then expressed in terms of the LT of the PDF of the interference, which is obtained later to evaluate \eqref{eq:Outage} and  \eqref{eq:Thr_Gen}\footnote{Expressions for the bit error probability can be derived by using the obtained SINR in the next section and following \cite{The2014Renzo,AlAmmouri2015Inband}.}.

\section{Performance Analysis}
The received signal is first convolved with the conjugated time-reversed pulse shape template, passed through a low-pass filter, and sampled at $t=t_o$. The baseband signal after filtering and sampling at the input of the decoder is given by:

\small
\begin{align}
y^{(i)}_{v}(t_o)=&y^{(i)}_{v}(t).* h^{(i)}_{v}(t-t_0)|_{t=t_o} \notag \\
=&A \sqrt{P^{(i)}_{v_o} r_o^{-\eta_{vv}} h_o} \mathcal{I}_{v} (\alpha_i,\alpha_i)+ \sum\limits_{k=1}^{K} \sum_{j \in \tilde{\Psi}^{(k)}_v} \Gamma^{(k)}_{v_{j}} \sqrt{P^{(k)}_{v_j} h^{(k)}_{v_j} \left(r^{(k)}_{v_j}\right)^{-\eta_{vv}}} \mathcal{I}_{v}(\alpha_i,\alpha_k) +\notag\\ &\sum\limits_{k=1}^{K} \sum_{j \in \tilde{\Psi}^{(k)}_{\bar{v}}} \Gamma^{(k)}_{{\bar{v}}_{j}} \sqrt{P^{(k)}_{{\bar{v}}_j} h^{(k)}_{\bar{v}_j} \left(r^{(k)}_{\bar{v}_j}\right)^{-\eta_{\bar{v}v}}}    \mathcal{C}_{v}(\alpha_i,\alpha_k)  + 
 \mathfrak{I}^{(i)}_{{\rm s}_{v}}(t) .* h^{(i)}_{v}(t-t_0)|_{t=t_o} +\sqrt{N_o |\mathcal{I}_{v} (\alpha_i,\alpha_i)|^2}.
\label{eq:base_band_matched}
\end{align}
\normalsize
where $h^{(i)}_{v}(t)$ is the combined matched and low-pass filter impulse response for a transceiver in the $i^{th}$ tier. The frequency domain representation $h^{(i)}_{v}(t)$ is given by

\small
\begin{align}
 H^{(i)}_{v}(f)=\left\{
	\begin{array}{ll}
		S^{*}(f,B_{v}(\alpha_i),b^{(i)}_{v}) \ \ \  \ \ \ \ \    -\frac{B_{v}(\alpha_i)}{2} &\leq f \leq \frac{B_{v}(\alpha_i)}{2}.\\
		0 & \! \!\! \!\! \! \rm{elsewhere}.
	\end{array}
\right.
 \label{matched2}
\end{align}\normalsize
where $S(f,B_{v}(\alpha_i),b^{(i)}_{v})$ represents the used pulse shape as discussed in section \ref{sec:PulseShaping}. 

The factors $\mathcal{I}(\cdot,\cdot)$ and $\mathcal{C}(\cdot,\cdot)$ in \eqref{eq:base_band_matched} represent the intra-mode (i.e., from UL-UL or DL-DL ) and cross-mode (i.e., from UL-DL or vice versa) effective received energy factors, respectively. From \eqref{inter1}, \eqref{inter2}, \eqref{matched2}, and expressing the convolution operation in the frequency domain, the pulse shaping and filtering factors are obtained as,

\small
\begin{equation} \label{fac1}
\mathcal{I}_{v} (\alpha_i,\alpha_k)=\int\limits_{-B_{v}(\alpha_i)/2}^{B_{v}(\alpha_i)/2} S^{*}(f,B_{v}(\alpha_i),b^{(i)}_{v})S(f-f^{(k)}_{v}+f^{(i)}_{v},B_{v}(\alpha_k),b^{(k)}_{v})df ,
\end{equation}
\begin{equation}  \label{fac2}
\mathcal{C}_{v} (\alpha_i,\alpha_k)=\int\limits_{-B_{v}(\alpha_i)/2}^{B_{v}(\alpha_i)/2} S^{*}(f,B_{v}(\alpha_i),b^{(i)}_{v})S(f-f^{(k)}_{\bar{v}}+f^{(i)}_{v},B_{\bar{v}}(\alpha_k),b^{(k)}_{\bar{v}})df ,
\end{equation}

\normalsize
It should be noted that although same mode links use similar pulse shapes in the same tier, the effective energy received from intra-mode intra-tier transmitters is not unity as shown in \eqref{fac1}. This is because \eqref{matched2} includes the combined impulse response of the matched and low-pass filters, which extracts the desired frequency range from the received signal. Consequently, the energy outside the desired BW is discarded and the energy contained within the pulse shape is no longer unity. Also, the cross-mode interference factor in \eqref{fac2} is strictly less than unity due to low-pass filtering, the possibly of different pulse shapes, and the partial overlap between cross-mode channels.

Let $ \Xi = \left\{r_o, r^{(i)}_{v_j}, h_o, h^{(i)}_{v_j}, P^{(i)}_{v_o}, P^{(i)}_{v_j}, h_s ; \forall i=\{1,...,K\}, v\in \{{\rm u,d} \}\right\}$, then conditioning on $\Xi$  the $\rm{SINR}$ is given by

\small
\begin{align}\label{eq:SINR}
&\!\!\!\!\!\!\!\!\!\!\!\!{\rm SINR}^{(i)}_{v}\left( \Xi\right) =\notag \\
&\!\!\!\!\!\!\!\!\!\!\!\!\frac{P^{(i)}_{v_o} r_o^{-\eta_{vv}} h_o}{\sum\limits_{k=1}^{K} \sum\limits_{j \in \tilde{\Psi}^{(k)}_v}  P^{(k)}_{v_j} h^{(k)}_{v_j} \left(r^{(k)}_{v_j}\right)^{-\eta_{vv}}  |\tilde{\mathcal{I}}_{v}(\alpha_i,\alpha_k)|^2 +\sum\limits_{k=1}^{K} \sum\limits_{j \in \tilde{\Psi}^{(k)}_{\bar{v}}}  P^{(k)}_{{\bar{v}}_j} h^{(k)}_{\bar{v}_j} \left(r^{(k)}_{\bar{v}_j}\right)^{-\eta_{\bar{v}v}} |\tilde{\mathcal{C}}_{v}(\alpha_i,\alpha_k)|^2 + \tilde{\sigma}_{{\rm s}_v}^2(\alpha_i) +N_o},
\end{align}\normalsize
where,\small
\begin{align}
|\tilde{\mathcal{I}}_{ v}(\alpha_i,\alpha_k)|^2&=\frac{|\mathcal{I}_{ v}(\alpha_i,\alpha_k)|^2}{| \mathcal{I}_{v}(\alpha_i,\alpha_i)|^2}, \\
|\tilde{\mathcal{C}}_{ v}(\alpha_i,\alpha_k)|^2&=\frac{|\mathcal{C}_{ v}(\alpha_i,\alpha_k)|^2}{| \mathcal{I}_{v}(\alpha_i,\alpha_i)|^2},
\end{align}\normalsize
and $ \tilde{\sigma}_{{\rm s}_v}^2(\cdot)$ is the residual SI power normalized by $| \mathcal{I}_{v}(\alpha_i,\alpha_i)|^2$. From \eqref{eq:SIu1} and \eqref{eq:SId1}, $ \tilde{\sigma}_{{\rm s}_v}^2$ can be expressed for the UL and DL as
\small
\begin{align}\label{eq:SIu2}
& \tilde{\sigma}_{{\rm s_{\rm u}}}^2(\alpha_i)= \beta^{(i)}_{\rm u} h_s P^{(i)}_{\rm d} |\tilde{\mathcal{C}}_{\rm u}(\alpha_i,\alpha_i)|^2.
\end{align}
\begin{align}\label{eq:SId2}
&\tilde{\sigma}_{{\rm s_{\rm d}}}^2(\alpha_i)= \left\{
	\begin{array}{ll}
		\beta_{\rm d} h_s P_{{\rm u}_{o}} |\tilde{\mathcal{C}}_{\rm d}(\alpha_i,\alpha_i)|^2.  \ \ \   & {\rm 2NT}  \\
		0. & {\rm 3NT}
	\end{array}
\right.
\end{align}\normalsize

The SINR in \eqref{eq:SINR} is used in the next section to evaluate the outage provability and rate as discussed in Section~\ref{method}.

\subsection{Performance Metrics}

\normalsize
From \eqref{eq:Outage} and \eqref{eq:SINR}, the outage probability in the link $v \in \{u,d\}$ in the $i^{th}$ tier can be written as,

\small
\begin{align}\label{eq:outageGeneral2}
\mathcal{O}_{v}^{(i)}(\theta)&=\mathbb{P} \left\{ \frac{P^{(i)}_{v_o} r_o^{-\eta_{vv}} h_o}{\sum\limits_{k=1}^{K} \mathfrak{I}_{v \rightarrow v}^{(k,i)}  |\tilde{\mathcal{I}}_{v}(\alpha_i,\alpha_k)|^2 + \sum\limits_{k=1}^{K} \mathfrak{I}_{\bar{v}\rightarrow v}^{(k,i)} |\tilde{\mathcal{C}}_{v }(\alpha_i,\alpha_k)|^2 + \tilde{\sigma}_{{\rm s}_v}^2(\alpha_i) +N_o}<\theta\right\}.
\end{align}\normalsize
where in general, $\mathfrak{I}_{v\rightarrow w}^{(k,i)}=\sum\limits_{j \in \tilde{\Psi}^{(k)}_v}  P^{(k)}_{v_j} h^{(k)}_{v_j} \left(r^{(k)}_{v_j}\right)^{-\eta_{vw}}$. By exploiting the exponential distribution of $h_o$, it can be written as

\small
\begin{align}\label{eq:outageGeneral3}
\!\!\!\!\!\!\!\!\!\!\!\!\!\!\!\! \mathcal{O}_{v}^{(i)}(\theta)&=1-\notag \\
&\mathbb{E} \left[ e^{\frac{-N_o r_o^{\eta_{vv}}\theta}{P^{(i)}_{v_o}}} e^{\frac{-\tilde{\sigma}_{{\rm s}_v}^2(\alpha_i) r_o^{\eta_{vv}}\theta}{P^{(i)}_{v_o}}}  \prod\limits_{k=1}^{K} \mathcal{L}_{\mathfrak{I}_{v \rightarrow v}^{(k,i)}} \left( \frac{ r_o^{\eta_{vv}}\theta  |\tilde{\mathcal{I}}_{v}(\alpha_i,\alpha_k)|^2 }{P^{(i)}_{v_o}} \right) \mathcal{L}_{\mathfrak{I}_{\bar{v}\rightarrow v}^{(k,i)}} \left( \frac{ r_o^{\eta_{vv}}\theta  |\tilde{\mathcal{C}}_{v}(\alpha_i,\alpha_k)|^2 }{P^{(i)}_{v_o}} \right)  \right].
\end{align}\normalsize
where the expectation is over $\{r_o,P^{(i)}_{v_o},\tilde{\sigma}_{{\rm s}_v}^2 \}$. Since the $\{r_o,P^{(i)}_{v_o}\}$ depends on the UEs type (CCU or CEU) as discussed in section \ref{sec:NetModel}, we present an explicit study for each type. The serving distances $r_o$ for CCUs and CEU s are characterized via the following lemma.

\begin{lemma}
The serving distance distribution for a randomly selected CCU or CEU given that it is connected to the $i^{th}$ tier denoted by $f_{R^{(i)}_c}(.)$ and $f_{R^{(i)}_e}(.)$, respectively, are given by the following equations,
\small
\begin{align}\label{eq:Dist1}
f_{R^{(i)}_c}(r)&= \frac{2 \pi \bar{\lambda_i} r \exp \left(-\pi \bar{\lambda_i} r^2\right)}{1-\exp \left(-\pi \bar{\lambda_i} \left(\frac{P_{\rm u}}{\rho^{(i)}}\right)^{\frac{2}{\eta_{\rm dd}}}\right)} \mathbbm{1}_{\left\{0\leq r \leq \left(\frac{P_{\rm u}}{\rho^{(i)}}\right)^{\frac{1}{\eta_{\rm dd}}}\right\}}(r),
\end{align}
\begin{align}\label{eq:Dist2}
f_{R^{(i)}_e}(r)&= 2 \pi \bar{\lambda_i} r \exp \left(-\pi \bar{\lambda_i} r^2+\pi \bar{\lambda_i} \left(\frac{P_{\rm u}}{\rho^{(i)}}\right)^{\frac{2}{\eta_{\rm dd}}}\right) \mathbbm{1}_{\left\{\left(\frac{P_{\rm u}}{\rho^{(i)}}\right)^{\frac{1}{\eta_{\rm dd}}} < r < \infty\right\}}(r).
\end{align}\normalsize
where $\bar{\lambda_i}= \sum\limits_{k=1}^{K} \frac{\tau_k^2}{\tau_k^2} \lambda_k$.
\begin{proof}
Refer to Appendix A.
\end{proof}
\end{lemma}

From Lemma 1, it is straightforward to find the probability that a randomly selected UE from the $i^{th}$ tier is a CCU or a CEU, which are given by,
\small
\begin{align}
\mathbb{P} \left\{ {\rm CCU} \right\}&=1-\exp \left(-\pi \bar{\lambda_i} \left(\frac{P_{\rm u}}{\rho^{(i)}}\right)^{\frac{2}{\eta_{\rm dd}}}\right) , \\
\mathbb{P} \left\{ {\rm CEU} \right\}&=\exp \left(-\pi \bar{\lambda_i} \left(\frac{P_{\rm u}}{\rho^{(i)}}\right)^{\frac{2}{\eta_{\rm dd}}}\right).
\end{align}\normalsize

By the law of total probability, the average outage probability can be expressed via the CCUs' outage probability and the CEUs' outage probability, denoted by $\mathcal{O}_{v^c}^{(i)}$ and $\mathcal{O}_{v^e}^{(i)}$, respectively, as
\small
\begin{align}\label{eq:TotalOutage}
\bar{\mathcal{O}}_{v}^{(i)}(\theta)=\mathcal{O}_{v^c}^{(i)}(\theta)\mathbb{P} \left\{ {\rm CCU} \right\}+\mathcal{O}_{v^e}^{(i)}(\theta)\mathbb{P} \left\{ {\rm CEU} \right\},
\end{align}\normalsize
where each of $\mathcal{O}_{v^c}^{(i)}(\theta)$ and $\mathcal{O}_{v^e}^{(i)}(\theta)$ is represented as in \eqref{eq:outageGeneral3}, but with the specific parameters related to the CCUs and the CEUs.  From \eqref{eq:outageGeneral3}, it is clear that the  LT of the aggregate interference from each tier $\mathfrak{I}_{w \rightarrow v}^{(k,i)}$ is required to evaluate $\mathcal{O}_{v^c}^{(i)}(\theta)$ and $\mathcal{O}_{v^e}^{(i)}(\theta)$. The aggregate interference, and hence its LT, depends on the spatial distribution of the set of interfering BSs and UEs in the tier, $\tilde{\Psi}^{(k)}_{\rm d}$ and $\tilde{\Psi}^{(k)}_{\rm u}$, respectively. The set of interfering BSs $\tilde{\Psi}^{(k)}_{\rm d}$ in the $k^{th}$ tier is the same as the original set of BSs ${\Psi}^{(k)}_{\rm d}$ excluding the transmitting BS itself in the DL and the serving BS in the UL. Hence, $\tilde{\Psi}^{(k)}_{\rm d}$ is a PPP with intensity $\lambda_k$. From UEs associations and $\lambda_{\rm u} \gg \sum_{k=1}^K \lambda_k$, the intensity of the interfering UEs $\tilde{\Psi}^{(k)}_{\rm u}$ on a certain channel in the $k^{th}$ tier is also $\lambda_k$. However, $\tilde{\Psi}^{(k)}_{\rm u}$ is not a PPP because only one UE can use a channel in each Voronoi-cell, which impose correlations among the positions of the interfering UEs on each channel and violates the PPP assumption. Furthermore, the employed association makes the set of interfering UEs  $\tilde{\Psi}^{(k)}_{\rm u}$ and the set of interfering BSs $\tilde{\Psi}^{(k)}_{\rm d}$ correlated. The inter-correlations between the interfering UEs and the cross-correlations between the UEs and BSs impede the model tractability. Hence, to maintain the tractability, we ignore these correlations. The used assumptions to keep the model tractability are formally stated below.

\begin{assumption}
 The set of interfering UEs $\tilde{\Psi}^{(k)}_{\rm u}$ in the $k^{th}$ tier is a PPP with intensity  $\lambda_k$.
\end{assumption}

\begin{assumption}
The point process $\tilde{\Psi}^{(k)}_{\rm d}$ for the interfering BSs and the point process $\tilde{\Psi}^{(k)}_{\rm u}$ for the interfering UEs both in the $k^{th}$ tier are independent.
\end{assumption}

\begin{assumption}
The point processes $\tilde{\Psi}^{(k)}_{\rm u}$'s which represent the interfering UEs connected to different tiers are independent from each other.
\end{assumption}

\begin{remark}
The previous assumptions are necessary to maintain the model tractability. Assumption 1 has been used and validated in \cite{On2014ElSawy, Hybrid2015Lee, Load2014AlAmmouri,AlAmmouri2015Inband}, Assumption 2 in \cite{AlAmmouri2015Inband,Throughput2016Goyal}, and Assumption 3 in \cite{On2014ElSawy}. It is important to mention that these assumptions ignore the mutual correlations between the interfering sources, however, the correlations between the interfering sources and the test receiver are captured through the proper calculation for the interference exclusion region enforced by association and/or UL power control. The accuracy of the developed model with Assumptions 1-3 is validated via independent Monte Carlo simulation in Section IV.
\end{remark}

Based on  Assumptions 1-3, the LT of the aggregated interference is always generated from a PPP $\Phi$, but with different parameters such as interference exclusion regions, interferers intensity, and transmit power distribution. For brevity, we present the following unified lemma for the LT of the aggregate interference generated from a homogeneous PPP with general parameters and use it to obtain all LTs in (19). 

\begin{lemma}
Let $\mathcal{L}_{\mathfrak{I}}(s)$ be the LT of the aggregate interference $\mathfrak{I}$ generated from a PPP network with intensity $\lambda$, i.i.d transmit powers $P_j$, unit means i.i.d exponentially distributed channel power gains $h_j$, and per-interferer protection region of $\mathcal{B}(o,a_j)$, where $\mathcal{B}(o,a_j)$ is a ball centered at the origin $(o)$ and has a radius $a_j$.
Then, $\mathcal{L}_{\mathfrak{I}}(s)$ is given by, 

\small
\begin{align}\label{eq:LTgeneral}
\mathcal{L}_{\mathfrak{I}}(s)=\exp \left( \frac{-2 \pi \lambda}{\eta-2} \mathbb{E}_{P} \left[a^{2-\eta} s P \  {}_2 F_1 \left[1,1-\frac{2}{\eta};2-\frac{2}{\eta}; -a^{-\eta} P s \right]  \right] \right),
\end{align}\normalsize
where $ {}_2 F_1(\cdot,\cdot;\cdot;\cdot)$ is the hyper-geometric function \cite{Handbook1964Abramowitz}, $\mathbb{E}_{P} [ \cdot ]$ is the expectation over the transmitted power of the sources, and $\eta>2$ is a general path loss exponent. For the special case of $a=0$, equation \eqref{eq:LTgeneral} reduces to,

\small
\begin{align}\label{eq:LTa=0}
\mathcal{L}_{\mathfrak{I}}(s)=\exp \left( -\frac{2 \pi^2 \lambda}{\eta} \mathbb{E}_{P} \left[ \left(s P \right)^{\frac{2}{\eta}} \right]  \csc \left(\frac{2}{\eta}  \right)   \right).
\end{align}\normalsize
\begin{proof}
Refer to Appendix B.
\end{proof}
\end{lemma}

Due to the expectation over power distribution, the LT expression in \eqref{eq:LTgeneral} has an integral over hyper-geometric function. If the interference exclusion distance $a$ around the test receiver is independent of the transmit powers, then the expression given by \eqref{eq:LTgeneral} can be lower-bounded by the simplified closed-form expression given in the following lemma.

\begin{lemma} \label{lem:LTgeneralapp}
Let $\mathcal{L}_{\mathfrak{I}}(s)$ be the LT of the aggregate interference generated from a PPP network with intensity $\lambda$, i.i.d transmit powers $P_j$, unit means i.i.d exponentially distributed channel power gains, and interference protection region of $\mathcal{B}(o,a)$, where $\mathcal{B}(o,a)$ is a ball centered at the origin $(o)$ and has a radius $a$. Assuming that $a$ is independent from $P_j$, $\forall j$, then $\mathcal{L}_{\mathfrak{I}}(s)$ can be lower-bounded by, 

\small
\begin{align}\label{eq:LTgeneralapp}
\mathcal{L}_{\mathfrak{I}}(s) \geq \exp \left( \frac{-2 \pi \lambda}{\eta-2} a^{2-\eta} s \mathbb{E} \left[P\right] \  {}_2 F_1 \left[1,1-\frac{2}{\eta};2-\frac{2}{\eta}; -a^{-\eta}\mathbb{E} \left[P\right]s \right]  \right).
\end{align}\normalsize
\begin{proof}
Refer to Appendix C.
\end{proof}
\end{lemma}

Lemma \ref{lem:LTgeneralapp} precludes the necessity to integrate over the PDF of the transmit power of the interfering sources and give the LT is a closed-form containing the first moment of the transmit power, which reduces the computational complexity of the LTs. For the sake of simplified expressions, we always use the bound in \eqref{eq:LTgeneralapp} whenever applicable and is verified in the section-IV. Using Lemma 2 and Lemma 3 , the LTs of the aggregated interference $\mathfrak{I}_{v \rightarrow v}^{(k,i)}$ for the UL and DL for CCUs and CEUs are given by the following lemma.

\begin{lemma}
Let $\mathcal{L}^{(c)}_{I^{(k,i)}_{{\rm u}\rightarrow {\rm u}}}$ $\left(\mathcal{L}^{(e)}_{I^{(k,i)}_{{\rm u}\rightarrow {\rm u}}}\right)$, $\mathcal{L}^{(c)}_{I^{(k,i)}_{{\rm d}\rightarrow {\rm u}}}$ $\left(\mathcal{L}^{(e)}_{I^{(k,i)}_{{\rm d}\rightarrow {\rm u}}}\right)$, $\mathcal{L}^{(c)}_{I^{(k,i)}_{{\rm d}\rightarrow {\rm d}}}$ $\left(\mathcal{L}^{(e)}_{I^{(k,i)}_{{\rm d}\rightarrow {\rm d}}}\right)$ and $\mathcal{L}^{(c)}_{I^{(k,i)}_{{\rm u}\rightarrow {\rm d}}}$ $\left(\mathcal{L}^{(e)}_{I^{(k,i)}_{{\rm u}\rightarrow {\rm d}}}\right)$ represent the LTs of the UL to UL, DL to UL, DL to DL, and UL to DL aggregate interference generated from the $k^{th}$ tier affecting a CCU (CEU) and its serving BS given that both of them are in the $i^{th}$ tier, then these LTs are given by

\small
\begin{align}
\!\!\!\!\!\!\!\!\!\!\!\!\!\!    \mathcal{L}^{(c)}_{I^{(k,i)}_{{\rm u}\rightarrow {\rm u}}}(s) &=  \exp \left( \frac{-2 \pi \lambda_k \left(\rho^{(k)}\right)^{1-\frac{2}{\eta_{\rm uu}}}}{\eta_{\rm uu}-2}  \mathbb{E}\left[ \left( P^{(k)}_{\rm u} \right)^{\frac{2}{\eta_{\rm uu}}} \right]  s \  {}_2 F_1 \left[1,1-\frac{2}{\eta_{\rm uu}},2-\frac{2}{\eta_{\rm uu}}, -\rho^{(k)} s \right]\right) ,
\end{align}
\begin{align}
\!\!\!\!\!\!\!\!\!\!\!\!\!\!   \mathcal{L}^{(c)}_{I^{(k,i)}_{{\rm d} \rightarrow {\rm u}}}(s) = \mathcal{L}^{(e)}_{I^{(k,i)}_{{\rm d}\rightarrow {\rm u}}}(s) =  \exp \left( -\frac{2 \pi^2 \lambda_k}{\eta_{\rm du}}  \left(s   P^{(k)}_{\rm d}  \right)^{\frac{2}{\eta_{\rm du}}}   \csc \left(\frac{2 \pi}{\eta_{\rm du}}  \right)\right),
\end{align}
\begin{align}
\!\!\!\!\!\!\!\!\!\!\!\!\!\!   \mathcal{L}^{(e)}_{I^{(k,i)}_{{\rm u}\rightarrow {\rm u}}}(s|r_o) & \approx   \exp \left( \frac{-2 \pi \lambda_k }{\eta_{\rm uu}-2}  \mathbb{E}_{P^{(k)}_{\rm u}}\left[ P^{(k)}_{\rm u}s  \  {}_2 F_1 \left[1,1-\frac{2}{\eta_{\rm uu}},2-\frac{2}{\eta_{\rm uu}}, - P^{(k)}_{\rm u} s r_o^{-\eta_{\rm uu}} \right]\right] \right),\\
& \gtrsim   \exp \left( \frac{-2 \pi \lambda_k }{\eta_{\rm uu}-2}  \mathbb{E}\left[ P^{(k)}_{\rm u}\right] s  \  {}_2 F_1 \left[1,1-\frac{2}{\eta_{\rm uu}},2-\frac{2}{\eta_{\rm uu}}, - \mathbb{E}\left[ P^{(k)}_{\rm u}\right] s r_o^{-\eta_{\rm uu}} \right]\right),
\end{align}
\begin{align}
\!\!\!\!\!\!\!\!\!\!\!\!\!\!  \mathcal{L}^{(c)}_{I^{(k,i)}_{{\rm d}\rightarrow {\rm d}}}(s|r_o)&=  \mathcal{L}^{(e)}_{I^{(k,i)}_{{\rm d}\rightarrow {\rm d}}}(s)  = \exp \left( \frac{-2 \pi \lambda_k}{\eta_{\rm dd}-2} \left(\frac{r_o \tau_i}{\tau_j}\right) ^{2-\eta_{\rm dd}}  s P^{(k)}_{{\rm d}} \  {}_2 F_1 \left[1,1-\frac{2}{\eta_{\rm dd}},2-\frac{2}{\eta_{\rm dd}}, -\left(\frac{r_o \tau_i}{\tau_j}\right)^{-\eta_{\rm dd}} P^{(k)}_{\rm d}  s \right]  \right),
\end{align}
\begin{align}\label{eq:IudCCU}
\!\!\!\!\!\!\!\!\!\!\!\!\!\!  \mathcal{L}^{(c)}_{I^{(k,i)}_{{\rm u}\rightarrow {\rm d}}}(s)&\approx   \exp \left( \frac{-2 \pi \lambda_k \left(\rho^{(k)}\right)^{1-\frac{2}{\eta_{\rm ud}}}}{\eta_{\rm ud}-2}  \mathbb{E}\left[ \left( P^{(k)}_{\rm u} \right)^{\frac{2}{\eta_{\rm ud}}} \right]  s \  {}_2 F_1 \left[1,1-\frac{2}{\eta_{\rm ud}},2-\frac{2}{\eta_{\rm ud}}, -\rho^{(k)} s \right]\right)U^{(k,i)}_1(r_o,s),
\end{align}
\begin{align}\label{eq:IudCEU}
\!\!\!\!\!\!\!\!\!\!\!\!\!\!   \mathcal{L}^{(e)}_{I^{(k,i)}_{{\rm u}\rightarrow {\rm d}}}(s) \approx  \exp \left( \frac{-2 \pi \lambda_k }{\eta_{\rm ud}-2}  \mathbb{E}\left[ P^{(k)}_{\rm u}\right] s  \  {}_2 F_1 \left[1,1-\frac{2}{\eta_{\rm ud}},2-\frac{2}{\eta_{\rm ud}}, - \mathbb{E}\left[ P^{(k)}_{\rm u}\right] s r_o^{-\eta_{\rm ud}} \right]\right)U^{(k,i)}_1(r_o,s),
\end{align}

\normalsize
where,
\small
\begin{align}
\!\!\!\!\!\!\!\!\!\!\!\!\!\! \mathbb{E}\left[ \left(P^{(k)}_{\rm u} \right)^{\zeta}\right]= \frac{\left(\rho^{(k)}\right)^{\zeta} \gamma \left(\frac{\zeta \eta_{\rm dd}}{2}+1, \pi \bar{\lambda}_k \left(\frac{P_{\rm u}}{\rho^{(k)}}\right)^{\frac{2}{\eta_{\rm dd}}}\right)}{\left( \pi \bar{\lambda}_k \right)^{\frac{\zeta \eta_{\rm dd}}{2}}}+\left(P_{\rm u}\right)^{\zeta} \exp \left(\pi \bar{\lambda}_k \left(\frac{P_{\rm u}}{\rho^{(k)}}\right)^{\frac{2}{\eta_{\rm dd}}} \right),
\end{align}
and $U^{(k,i)}_1(r_o,s)$ is the LT of intra-cell interference in the 3NT case, which is expressed as
\begin{align}\label{eq:U1EX}
&\!\!\!\!\!\!\!\!\!\!\!\!\!\! U^{(k,i)}_1(r_o,s)=\notag \\
&\!\!\!\!\!\!\!\!\!\!\!\!\!\!  \left\{
	\begin{array}{ll}
		\int\limits_{0}^{\left(\frac{P_{\rm u}}{\rho^{(i)}}\right)^{\frac{1}{\eta_{\rm ud}}}}\int\limits_{\delta_o}^{\pi} \frac{\mathbb{P} \left\{ {\rm CCU} \right\} f_{R^{(i)}_c}(r)}{1+  s  \rho^{(i)} \left(1+(\frac{r_o}{r})^2-2\frac{r_o}{r} \cos (\delta) \right)^{\frac{-\eta_{\rm ud}}{2}}} \frac{d\delta dr}{\pi-\delta_o}+ \int\limits_{\left(\frac{P_{\rm u}}{\rho^{(i)}}\right)^{\frac{1}{\eta_{\rm ud}}}}^{\infty}\int\limits_{\delta_o}^{\pi} \frac{\mathbb{P} \left\{ {\rm CEU} \right\} f_{R^{(i)}_e}(r)}{1+  s  P_{\rm u} \left(r^2+r_o^2-2 r_o r \cos (\delta) \right)^{\frac{-\eta_{\rm ud}}{2}}} \frac{d\delta dr}{\pi-\delta_o}.  \ \ \     &  \underset{{i=k}}{\rm 3NT} \\
		1.  \ \ \     & {\rm O.W}\\
	\end{array}
\right.
\end{align}\normalsize
and $\gamma(\cdot,\cdot)$ is the lower incomplete gamma function \cite{Handbook1964Abramowitz}.
\begin{proof}
Refer to Appendix D.
\end{proof}
\end{lemma}

Note that $U^{(k,i)}_1(\cdot,\cdot)$ in equations \eqref{eq:IudCCU} and \eqref{eq:IudCEU} represents the intra-cell interference, in which it has an effect on the 3NT case from tier that tagged transceiver belongs to. For the sake of simplified expressions, we also present a closed-form approximation for $U^{(k,i)}_1(\cdot,\cdot)$ in the following lemma.

\begin{lemma}
The LT of the intra-cell interference given in equation \eqref{eq:U1EX} can be approximated by

\small
\begin{align}\label{eq:U1App}
&\!\!\!\!\!\!\!\!\!\!\!\!\!\! U^{(k,i)}_1(r_o,s)=\notag \\
&\!\!\!\!\!\!\!\!\!\!\!\!\!\!  \left\{
	\begin{array}{ll}
		\frac{\mathbb{P} \left\{ {\rm CCU} \right\} }{1+  s  \rho^{(i)} \left(1+(\frac{r_o}{\bar{r}_c})^2+\frac{2 \sin (\delta_o)}{\pi-\delta_o} \frac{r_o}{\bar{r}_c} \right)^{\frac{-\eta_{\rm ud}}{2}}} +  \frac{\mathbb{P} \left\{ {\rm CEU} \right\} }{1+  s  P_{\rm u} \left(\bar{r}_e^2+r_o^2+\frac{2 \sin (\delta_o)}{\pi-\delta_o} r_o \bar{r}_e  \right)^{\frac{-\eta_{\rm ud}}{2}}}  .  \ \ \     &  \underset{{i=k}}{\rm 3NT} \\
		1.  \ \ \     & {\rm O.W}\\
	\end{array}
\right.
\end{align}\normalsize
where \small
\begin{align}
\bar{r}_c=\frac{{\rm erf} \left( \sqrt{\pi \bar{\lambda_i}} \left(\frac{P_{\rm u}}{\rho^{(i)}}\right)^{\frac{1}{\eta_{\rm ud}}} \right)-2 \sqrt{\bar{\lambda_i}} \left(\frac{P_{\rm u}}{\rho^{(i)}}\right)^{\frac{1}{\eta_{\rm ud}}} \exp \left(- \pi \bar{\lambda_i} \left(\frac{P_{\rm u}}{\rho^{(i)}}\right)^{\frac{2}{\eta_{\rm ud}}} \right)}{2 \sqrt{\bar{\lambda_i}}\left(1- \exp \left(- \pi \bar{\lambda_i} \left(\frac{P_{\rm u}}{\rho^{(i)}}\right)^{\frac{2}{\eta_{\rm ud}}} \right)\right)}.
\end{align}
\begin{align}
\bar{r}_e=\frac{\exp \left( \pi \bar{\lambda_i} \left(\frac{P_{\rm u}}{\rho^{(i)}}\right)^{\frac{2}{\eta_{\rm ud}}} \right) {\rm erfc} \left( \sqrt{\pi \bar{\lambda_i}} \left(\frac{P_{\rm u}}{\rho^{(i)}}\right)^{\frac{1}{\eta_{\rm ud}}} \right)-2 \sqrt{\bar{\lambda_i}} \left(\frac{P_{\rm u}}{\rho^{(i)}}\right)^{\frac{1}{\eta_{\rm ud}}} }{2 \sqrt{\bar{\lambda_i}}}.
\end{align}\normalsize
where ${\rm erf} (\cdot)$ and ${\rm erfc} (\cdot)$ are the error function and the complementary error function, respectively \cite{Handbook1964Abramowitz}. 
\begin{proof}
By substituting $r$ and $\cos (\delta)$ by their average values.  
\end{proof}
\end{lemma}

Using the results in Lemmas 1-5 along with (19), the outage probabilities for all types of connections in the depicted system model are characterized via the following theorem.

\begin{theorem}
The outage probabilities in the UL and the DL in the $i^{th}$ tier for CCUs and CEUs are given by,
\small
\begin{align}\label{eq:OutageUL1}
\!\!\!\!\!\!\!\!\!\!\!\!\! \mathcal{O}_{{\rm u}^c}^{(i)}(\theta)&=1-e^{\frac{-N_o \theta}{\rho^{(i)}}} U^{(i)}_{\rm SI_ u}\left(\frac{\theta}{\rho^{(i)}}\right) \prod\limits_{k=1}^{K} \mathcal{L}^{(c)}_{\mathfrak{I}_{\rm u \rightarrow u}^{(k,i)}} \left( \frac{ \theta  |\tilde{\mathcal{I}}_{{\rm u}}(\alpha_i,\alpha_k)|^2 }{\rho^{(i)}} \right) \mathcal{L}^{(c)}_{\mathfrak{I}_{{\rm d \rightarrow u}}^{(k,i)}} \left( \frac{ \theta  |\tilde{\mathcal{C}}_{{\rm u}}(\alpha_i,\alpha_k)|^2 }{\rho^{(i)}} \right),
\end{align}
\begin{align}\label{eq:OutageUL2}
\!\!\!\!\!\!\!\!\!\!\!\!\! \mathcal{O}_{{\rm u}^e}^{(i)}(\theta)&=1- \int\limits_{\left(\frac{P_{\rm u}}{\rho^{(i)}}\right)^{\frac{1}{\eta_{\rm uu}}}}^{\infty}  e^{\frac{-N_o r_o^{\eta_{\rm uu}}\theta}{P_{\rm u}}}U^{(i)}_{\rm SI_ u}\left(\frac{\theta r_o^{\eta_{\rm uu}}}{P_{\rm u}}\right) f_{R^{(i)}_e}(r_o) \notag \\
  & \ \ \ \ \ \ \ \ \ \ \ \ \ \ \ \ \ \ \times \prod\limits_{k=1}^{K} \mathcal{L}^{(e)}_{\mathfrak{I}_{\rm u \rightarrow u}^{(k,i)}} \left( \frac{ r_o^{\eta_{\rm uu}}\theta  |\tilde{\mathcal{I}}_{{\rm u}}(\alpha_i,\alpha_k)|^2 }{P_{\rm u}} \right) \mathcal{L}^{(e)}_{\mathfrak{I}_{{\rm d \rightarrow u}}^{(k,i)}} \left( \frac{ r_o^{\eta_{\rm uu}}\theta  |\tilde{\mathcal{C}}_{{\rm u}}(\alpha_i,\alpha_k)|^2 }{P_{\rm u}} \right) dr_o,
\end{align}
\begin{align}
\!\!\!\!\!\!\!\!\!\!\!\!\! \mathcal{O}_{{\rm d}^{c}}^{(i)}(\theta)&=1-  \int\limits_{0}^{\left(\frac{P_{\rm u}}{\rho^{(i)}}\right)^{\frac{1}{\eta_{\rm dd}}}} e^{\frac{-N_o r_o^{\eta_{\rm dd}}\theta}{P^{(i)}_{\rm d}}} U^{(i)}_{\rm SI_ d}\left(\frac{\theta \rho r_o^{2\eta_{\rm dd}}}{P_{\rm d}^{(i)}}\right) f_{R^{(i)}_c}(r_o) \notag \\
& \ \ \ \ \ \ \ \ \ \ \ \ \ \ \ \ \ \ \times \prod\limits_{k=1}^{K} \mathcal{L}^{(c)}_{\mathfrak{I}_{\rm d \rightarrow d}^{(k,i)}} \left( \frac{ r_o^{\eta_{\rm dd}}\theta  |\tilde{\mathcal{I}}_{{\rm d}}(\alpha_i,\alpha_k)|^2 }{P^{(i)}_{{\rm d}}} \right) \mathcal{L}^{(c)}_{\mathfrak{I}_{u \rightarrow d}^{(k,i)}} \left( \frac{ r_o^{\eta_{\rm dd}}\theta  |\tilde{\mathcal{C}}_{{\rm d}}(\alpha_i,\alpha_k)|^2 }{P^{(i)}_{{\rm d}}} \right) dr_o,
\end{align}
\begin{align}
\!\!\!\!\!\!\!\!\!\!\!\!\! \mathcal{O}_{{\rm d}^{e}}^{(i)}(\theta)&=1-\int\limits_{\left(\frac{P_{\rm u}}{\rho^{(i)}}\right)^{\frac{1}{\eta_{\rm dd}}}}^{\infty} e^{\frac{-N_o r_o^{\eta_{\rm dd}}\theta}{P^{(i)}_{\rm d}}} U^{(i)}_{\rm SI_ d}\left(\frac{\theta P_{\rm u} r_o^{\eta_{\rm dd}}}{P_{\rm d}^{(i)}}\right) f_{R^{(i)}_e}(r_o) \notag \\
& \ \ \ \ \ \ \ \ \ \ \ \ \ \ \ \ \ \ \times \prod\limits_{k=1}^{K} \mathcal{L}^{(e)}_{\mathfrak{I}_{\rm d \rightarrow d}^{(k,i)}} \left( \frac{r_o^{\eta_{\rm dd}}\theta  |\tilde{\mathcal{I}}_{{\rm d}}(\alpha_i,\alpha_k)|^2 }{P^{(i)}_{{\rm d}}} \right) \mathcal{L}^{(e)}_{\mathfrak{I}_{u\rightarrow d}^{(k,i)}} \left( \frac{ r_o^{\eta_{\rm dd}}\theta  |\tilde{\mathcal{C}}_{{\rm d}}(\alpha_i,\alpha_k)|^2 }{P^{(i)}_{{\rm d}}} \right) dr_o,
\end{align}\normalsize
where,\small
\begin{align}\label{eq:SIu}
\!\!\!\!\!\!\!\!\!\!&U^{(i)}_{\rm SI_ u}(x)=\int\limits_{0}^{\infty} \exp \left(- x\beta^{(i)}_{\rm u} h P^{(i)}_{\rm d} |\tilde{\mathcal{C}}_{\rm u}(\alpha_i,\alpha_i)|^2  \right) f_{H_s}(h) dh,
\end{align}
\begin{align}\label{eq:SId}
\!\!\!\!\!\!\!\!\!\!&U^{(i)}_{\rm SI_ d}(x)= \left\{
	\begin{array}{ll}
		\int\limits_{0}^{\infty} \exp \left(- x \beta_{\rm d} h  |\tilde{\mathcal{C}}_{\rm d}(\alpha_i,\alpha_i)|^2\right) f_{H_s}(h) dh.  \ \ \   & {\rm 2NT}  \\
		1. & {\rm 3NT}
	\end{array}
\right.
\end{align}\normalsize
and $f_{H_s}(\cdot)$ is the distribution of the SIC power. $f_{R^{(i)}_c}(\cdot)$ and $f_{R^{(i)}_e}(\cdot)$ are given in equations \eqref{eq:Dist1} and \eqref{eq:Dist2}, respectively, and the LTs are given in Lemma 4.
\begin{proof}
Refer to Appendix E.
\end{proof}
\end{theorem}

A special case of interest that leads to simple forms of the outage probability is presented in the following corollary.

\begin{corollary}
In an interference limited dense single tier cellular network with unbinding UL transmit power, the outage probability in the DL and UL, assuming $\eta_{\rm dd}=\eta_{\rm uu}=\eta_{\rm ud}=4$, $\eta_{\rm du}=3$, $\delta_o=90^o$, and $h_s \sim \exp(1)$, are given by

\small
\begin{align}\label{eq:OutageUL11}
\!\!\!\!\!\!\!\!\!\!\!\!\! \mathcal{O}_{{\rm u}}(\theta)& \approx 1-\frac{\exp \left( - \sqrt{ \theta } \arctan \left( \sqrt{ \theta }\right) -\frac{4 \pi^2 \lambda}{3\sqrt{3}}  \left(\frac{ \theta  |\tilde{\mathcal{C}}_{{\rm u}}(\alpha,\alpha)|^2 P_{\rm d}  }{\rho}     \right)^{\frac{2}{3}} \right)  }{1+\beta_{\rm u}  P_{\rm d} |\tilde{\mathcal{C}}_{\rm u}(\alpha,\alpha)|^2 \frac{\rho}{\theta}}
\end{align}

\begin{align}\label{eq:OutageDL11}
\!\!\!\!\!\!\!\!\!\!\!\!\! \mathcal{O}_{{\rm d}}(\theta)& \approx 1-\int\limits_{0}^{\infty} 2 \pi \lambda r_o U_{\rm NT}(\theta,r_o)  \notag \\
&\!\!\!\!\!\!\!\!\!\!\!\!\! \times    \exp \left(- \pi \lambda r_o^2 - \pi \lambda r_o^2 \sqrt{\theta} \arctan \left(\sqrt{\theta} \right)- \sqrt{\frac{\rho r_o^4 \theta|\tilde{\mathcal{C}}_{{\rm d}}(\alpha,\alpha)|^2 }   {P_{\rm d}}} \arctan \left( \sqrt{\frac{\rho r_o^4 \theta |\tilde{\mathcal{C}}_{{\rm d}}(\alpha,\alpha)|^2 }   {P_{\rm d}}} \right) \right)  dr_o
\end{align}\normalsize
where,\small
\begin{align}\label{eq:UNT}
\!\!\!\!\!\!\!\!\!\!U_{\rm NT}(\theta,r_o)&= \left\{
	\begin{array}{ll}
		\frac{P_{\rm d} }{P_{\rm d}+\beta_{\rm d} |\tilde{\mathcal{C}}_{{\rm d}}(\alpha,\alpha)|^2 \rho r_o^{8}\theta}.  \ \ \   & {\rm 2NT}  \\
		\int\limits_{0}^{\infty}\int\limits_{\delta_o}^{\pi} \frac{  P_{\rm d} \lambda r \exp \left(- \pi \lambda r^2 \right)}{P_{\rm d}+ r_o^4 \theta |\tilde{\mathcal{C}}_{{\rm d}}(\alpha,\alpha)|^2  \rho \left(1+(\frac{r_o}{r})^2-2\frac{r_o}{r} \cos (\delta) \right)^{-2}} d\delta dr. & {\rm 3NT}
	\end{array}
\right. \\
&\approx\left\{
	\begin{array}{ll}
		\frac{P_{\rm d} }{P_{\rm d}+\beta_{\rm d} |\tilde{\mathcal{C}}_{{\rm d}}(\alpha,\alpha)|^2 \rho r_o^{8}\theta}.  \ \ \   & {\rm 2NT}  \\
		 \frac{  P_{\rm d}}{P_{\rm d}+ r_o^4 \theta |\tilde{\mathcal{C}}_{{\rm d}}(\alpha,\alpha)|^2  \rho \left(1+4 \lambda  r_o^2+\frac{8}{\pi} \sqrt{\lambda}r_o  \right)^{-2}} dr. & {\rm 3NT}
	\end{array}
\right.
\end{align}\normalsize
\begin{proof}
The expressions follow from Theorem 1 and Lemmas 4-5 by considering a single tier network and setting $P_{\rm u} \rightarrow \infty$.
\end{proof}
\end{corollary}

Following \eqref{eq:Thr_Gen}, the rate can be expressed in terms of the outage probability as follows

\small
\begin{align}\label{eq:Thr_Out}
\mathcal{R} (\theta)={{\rm BW}} \log_2 \left(1+\theta \right) \left(1-\mathcal{O}(\theta) \right).
\end{align}\normalsize
Hence, general expressions for the $\alpha$-duplex rate in a multi-tier network can be obtained by directly substituting the outage probability expressions from Theorem 1 in equation \eqref{eq:Thr_Gen}. For the sake of brevity, we only list the rate expressions for a special case of interest in the following Corollary.

\begin{corollary}
In an interference limited dense single tier cellular network with unbinding UL transmit power, the average rate in the DL and UL, assuming $\eta_{dd}=\eta_{uu}=\eta_{ud}=4$, $\eta_{du}=3$, $\delta_o=90^o$, and $h_s \sim \exp(1)$, are given by

\small
\begin{align}\label{eq:RateUL11}
\!\!\!\!\!\!\!\!\!\!\!\!\! \mathcal{R}_{{\rm u}}(\theta)& \approx \frac{(B_{\rm u}+(\epsilon+1) \alpha B) \log_2\left(1+\theta \right)}{1+\beta_{\rm u}  P_{\rm d} |\tilde{\mathcal{C}}_{\rm u}(\alpha,\alpha)|^2 \frac{\rho}{\theta}}\exp \left( - \sqrt{ \theta } \arctan \left( \sqrt{ \theta }\right) -\frac{4 \pi^2 \lambda}{3\sqrt{3}}  \left(\frac{ \theta  |\tilde{\mathcal{C}}_{{\rm u}}(\alpha,\alpha)|^2 P_{\rm d}  }{\rho}     \right)^{\frac{2}{3}} \right)  .
\end{align}

\begin{align}\label{eq:RateDL11}
\!\!\!\!\!\!\!\!\!\!\!\!\! \mathcal{R}_{{\rm d}}(\theta)& \approx 2 \pi \lambda(B_{\rm d}+(\epsilon+1) \alpha B) \log_2\left(1+\theta \right) \notag \\
&\!\!\!\!\!\!\!\!\!\!\!\!\! \times \int\limits_{0}^{\infty}  r_o    \exp \left(- \pi \lambda r_o^2 - \pi \lambda r_o^2 \sqrt{\theta} \arctan \left(\sqrt{\theta} \right)- \sqrt{\frac{\rho r_o^4 \theta|\tilde{\mathcal{C}}_{{\rm d}}(\alpha,\alpha)|^2 }   {P_{\rm d}}} \arctan \left( \sqrt{\frac{\rho r_o^4 \theta |\tilde{\mathcal{C}}_{{\rm d}}(\alpha,\alpha)|^2 }   {P_{\rm d}}} \right) \right) U_{\rm NT}(\theta,r_o) dr_o
\end{align}\normalsize
where,\small
\begin{align}\label{eq:RUNT}
\!\!\!\!\!\!\!\!\!\!U_{\rm NT}(\theta,r_o)&= \left\{
	\begin{array}{ll}
		\frac{P_{\rm d} }{P_{\rm d}+\beta_{\rm d} |\tilde{\mathcal{C}}_{{\rm d}}(\alpha,\alpha)|^2 \rho r_o^{8}\theta}.  \ \ \   & {\rm 2NT}  \\
		\int\limits_{0}^{\infty}\int\limits_{\delta_o}^{\pi} \frac{  P_{\rm d} \lambda r \exp \left(- \pi \lambda r^2 \right)}{P_{\rm d}+ r_o^4 \theta |\tilde{\mathcal{C}}_{{\rm d}}(\alpha,\alpha)|^2  \rho \left(1+(\frac{r_o}{r})^2-2\frac{r_o}{r} \cos (\delta) \right)^{-2}} d\delta dr. & {\rm 3NT}
	\end{array}
\right. \\
&\approx\left\{
	\begin{array}{ll}
		\frac{P_{\rm d} }{P_{\rm d}+\beta_{\rm d} |\tilde{\mathcal{C}}_{{\rm d}}(\alpha,\alpha)|^2 \rho r_o^{8}\theta}.  \ \ \   & {\rm 2NT}  \\
		 \frac{  P_{\rm d}}{P_{\rm d}+ r_o^4 \theta |\tilde{\mathcal{C}}_{{\rm d}}(\alpha,\alpha)|^2  \rho \left(1+4 \lambda  r_o^2+\frac{8}{\pi} \sqrt{\lambda}r_o  \right)^{-2}} dr. & {\rm 3NT}
	\end{array}
\right.
\end{align}\normalsize
\begin{proof}
Follows from Corollary 1 and equation \eqref{eq:Thr_Out}.
\end{proof}
\end{corollary}

From the last corollary, we can find the critical SIC $\beta_{\rm d}$ at which the 2NT outperforms the 3NT as a function of the serving distance ($r_o$). This value is given by the following corollary.
\begin{corollary}
In an interference limited dense single tier cellular network with unbinding UL transmit power, the approximate minimum value of $\beta_{\rm d}$ required in the 2NT to outperform 3NT as a function of the serving distance ($r_o$), assuming $\eta_{ud}=4$, $\delta_o=90^o$, fully-overlapped channels ($\alpha=1$ and $|\tilde{\mathcal{C}}_{{\rm d}}(\alpha,\alpha)|^2=1$), and $h_s \sim \exp(1)$, is given by

\small
\begin{align}\label{eq:Cond1}
\beta_{\rm d} \approx \left(4 \lambda r_o^4+\frac{8}{\pi} \sqrt{\lambda} r_o^3 +r_o^2\right)^{-2}
\end{align}\normalsize
\begin{proof}
Follows from equation (54).
\end{proof}
\end{corollary}

Equation \eqref{eq:Cond1} expresses the critical value of $\beta_{\rm d}$ as a function of $r_o$ and $\lambda$. To get more insights on the critical value of $\beta_{\rm d}$ with respect to the BSs' intensity $\lambda$ , we assume that the tagged UE is located at the average serving distance, this assumption reduces \eqref{eq:Cond1} to

\small
\begin{align}\label{eq:Cond2}
\beta_{\rm d} \approx \frac{16 \lambda^2}{9}
\end{align}\normalsize

In the next section, Theorem 1, Lemmas 1-5, and Corollaries 1-3 are used to analyze the performance of cellular network under 2NT and 3NT operations.

\section{Simulations and Numerical Results}
\begin{table} []
\caption{\; Parameters Values.}
\centering
\begin{tabular}{|l|l|l|l|}
\hline
\rowcolor[HTML]{C0C0C0}
\textbf{Parameter} & \textbf{Value}      & \textbf{Parameter} & \textbf{Value}  \\ \hline
$P_{\rm{u}}$        &  3 W                & $P_{\rm{d}}$              & 5 W             \\ \hline
$\lambda$          & 1 $\text{BSs/km}^2$ & $N_o$              & 0 \\ \hline
$B^{\rm HD}_{\rm u}$, $B^{\rm HD}_{\rm d}$              & 1 MHz               &     $\theta$          & 1           \\ \hline
$\beta_{\rm d}$            &$-75$ dB              &    $\beta_{\rm u}$            & $-110$ dB      \\ \hline
$\rho$   & -60 dBm                 & $\epsilon$  & 0.03134               \\ \hline
$b_{\rm d}/b_{\rm u}$   & Sinc/Sinc$^2$                 & $\delta_o$   & $90^o$               \\ \hline
$\eta_{\rm uu}$, $\eta_{\rm dd}$, $\eta_{\rm ud}$   & 4                 & $\eta_{\rm du}$   & 3               \\ \hline
\end{tabular}
\label{TB:parameters}\vspace{-0.5cm}
\end{table}

Throughout this section, we verify the developed mathematical paradigm via independent system level simulations, where the BSs are realized via a PPP over an area of 600 ${\rm km}^2$. Then, the UEs are distributed uniformly over the area such that each BS has at least two UEs within its association area. Each BS randomly selects two UEs to serve such that the $\delta_o$ angle separation as illustrated in Fig. \ref{fig:Network4}  is satisfied. The SINR is calculated by summing the interference powers from all the UEs and the BSs after multiplying by the effective interference factors. In the UL, the transmit powers of the UEs are set according to the power control discussed in  Section II. The results are taken for the UE and the BSs that are closest to the origin to avoid the edge effect. Unless otherwise stated, the parameters values in Table 1 are used. Note that for the average SIC power, the maximum reported value according to \cite{Full2015Goyal} is -110 dB, and hence, we set $\beta_{\rm u}$ to -110 and consider that $\beta_{\rm d} \geq \beta_{\rm u}$ because the BSs are more likely to have more powerful SIC capabilities.

For the pulse shaping, we consider two basic pulse shapes, namely, Sinc$^2$ and Sinc pulse shape\footnote{Employing and designing more sophisticated pulse shapes for specific purposes is left for future work.}, which have the FTs given in \eqref{eq:PulseShapes},

\small
\begin{align}\label{eq:PulseShapes}
S(f,\rm{BW},b)=\left\{
	\begin{array}{ll}
		\frac{\rm{SINC(\frac{2 f}{\rm{BW}})}}{\sqrt{\int\limits_{- \infty}^{\infty} {\rm SINC}^2\left(\frac{2 f}{\rm{BW}}\right) df}}  & \mbox{ } b = \rm{Sinc.} \\
		\frac{\rm{SINC^2(\frac{2 f}{\rm{BW}})}}{\sqrt{\int\limits_{- \infty}^{\infty} {\rm SINC}^4\left(\frac{2 f}{\rm{BW}}\right) df}} & \mbox{ } b = \rm{Sinc^2 .}
	\end{array}
\right.
\end{align}\normalsize
where $b =$ Sinc when the Sinc pulse is considered and $b =$ Sinc$^2$ when Sinc$^2$ pulse is considered. Unless otherwise stated, the SIC power distribution $f_{H_s}(\cdot)$ is assumed to be exponentially distributed with unit mean.

\begin{figure*}[t!]
    \centering
    \begin{subfigure}[b]{0.3\textwidth}
        \centering
        \includegraphics[width=2.08in]{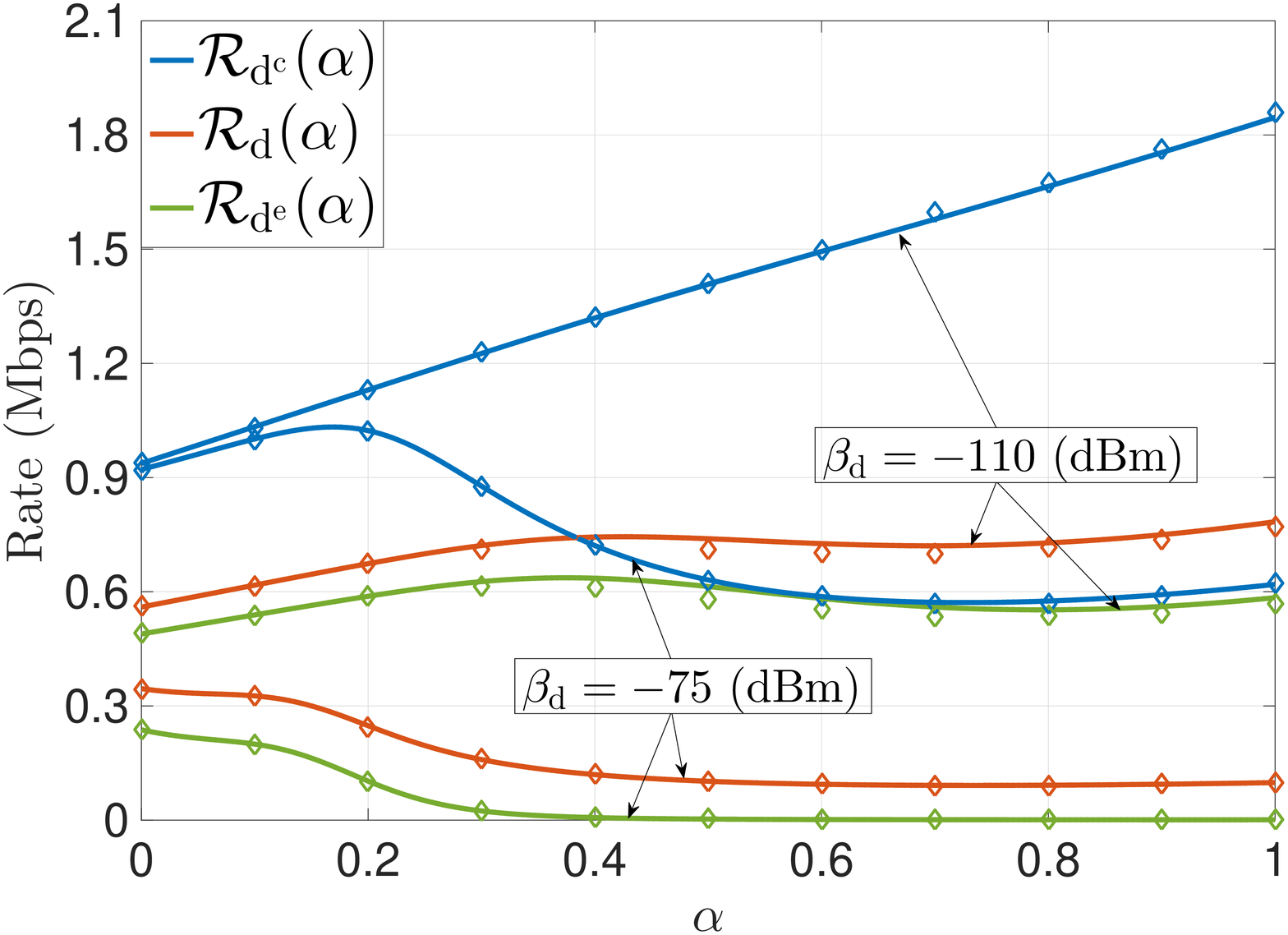}
        \caption{2NT DL throughput.}\label{fig:Ex1_1}
    \end{subfigure}%
    ~ 
    \begin{subfigure}[b]{0.3\textwidth}
        \centering
        \includegraphics[width=2.1in]{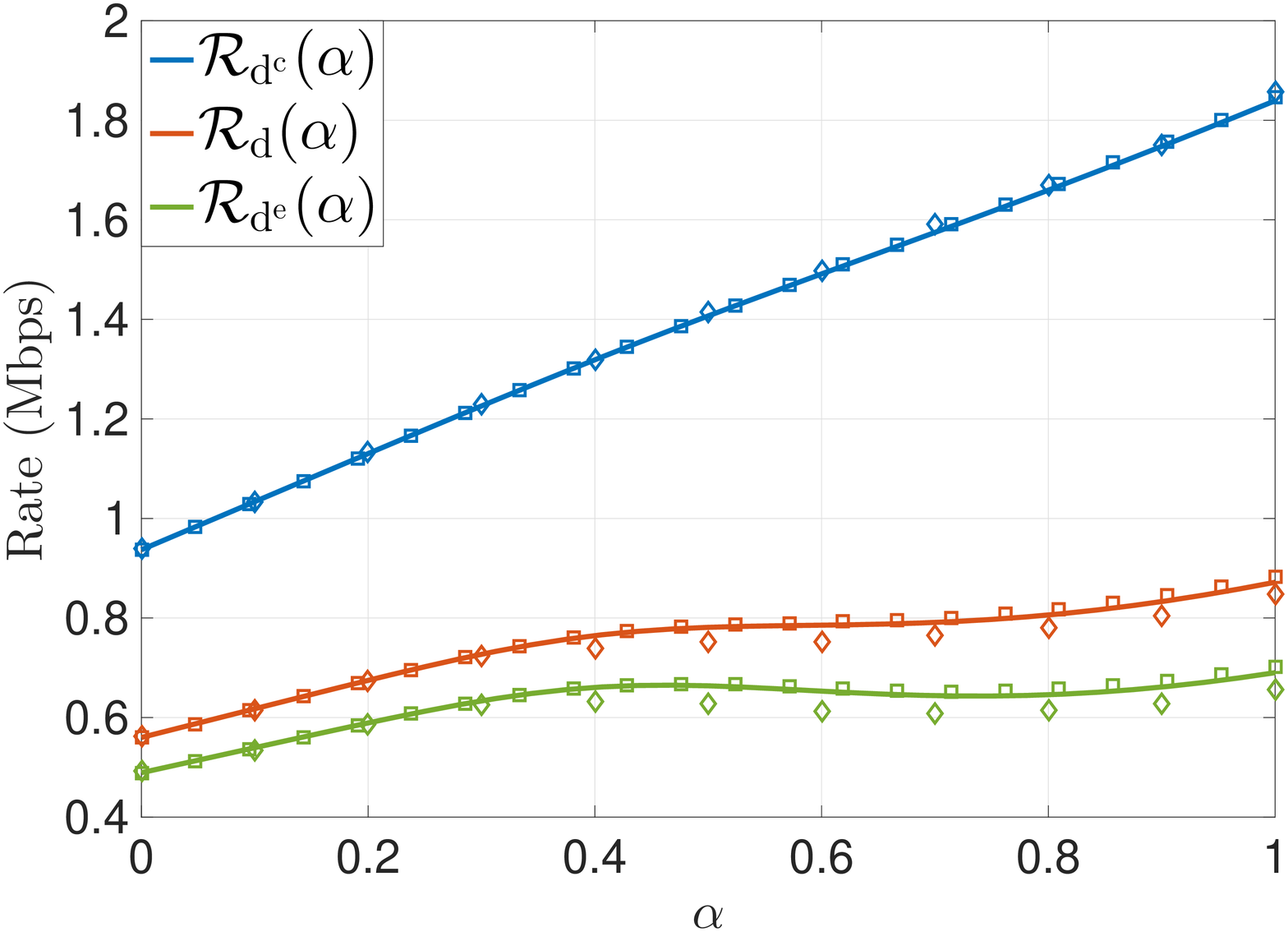}
        \caption{3NT DL throughput.}\label{fig:Ex1_2}
    \end{subfigure}
    ~ 
    \begin{subfigure}[b]{0.3\textwidth}
        \centering
        \includegraphics[width=2.1in]{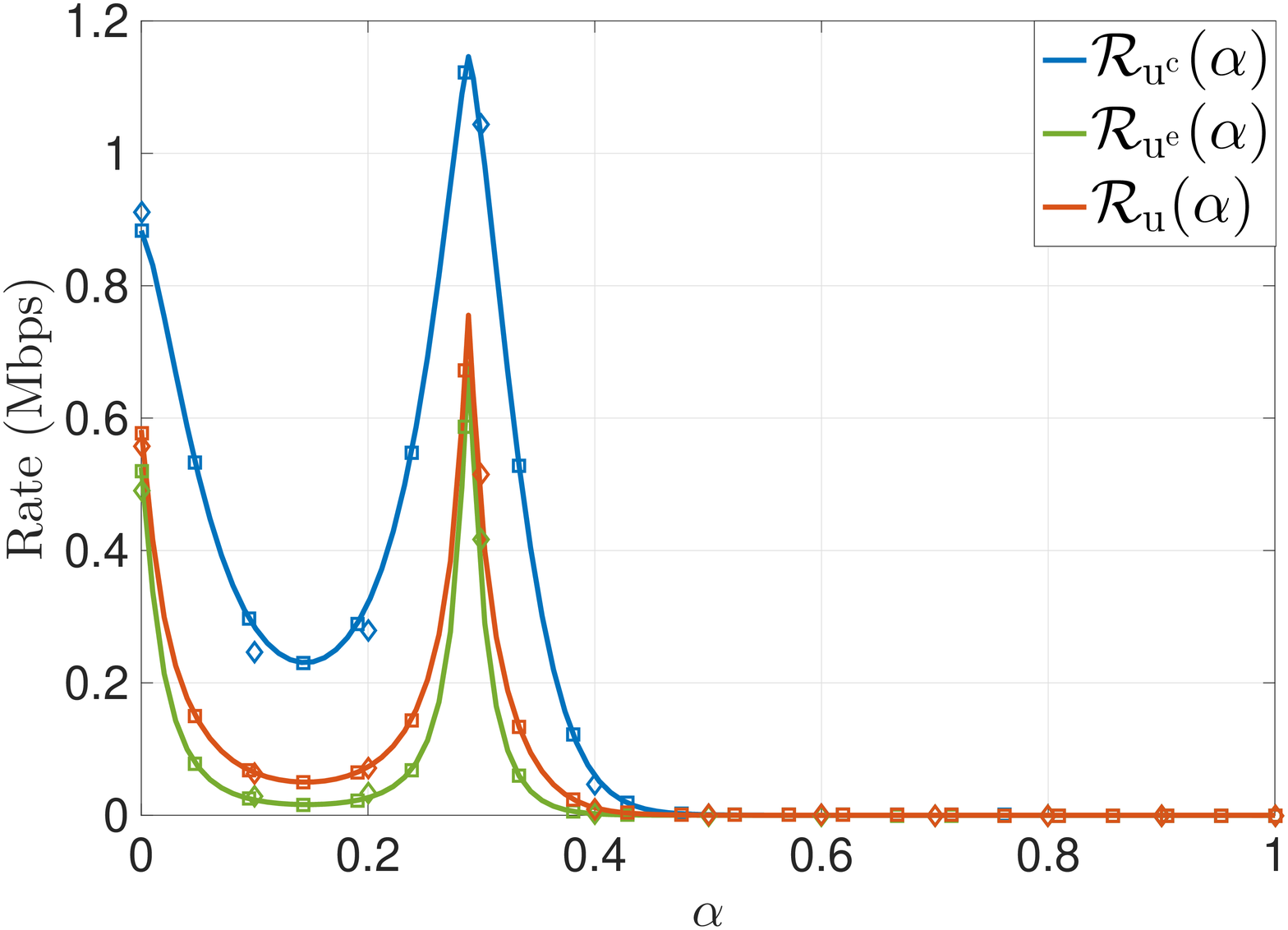}
        \caption{UL throughput.}\label{fig:Ex1_3}
    \end{subfigure}
    \caption{Rates vs $\alpha$ for the 3NT and 2NT.}\label{fig:Ex1}\vspace{-0.5cm}
\end{figure*}

Fig. \ref{fig:Ex1} shows the rate variation for UL and DL versus $\alpha$ for the 2NT and 3NT, where $\alpha=0$ and $\alpha=1$ represent the FD and the HD cases, respectively, the solid lines represent the analytical results obtained from Theorem 1 with the exact LTs in Lemma 4, diamonds represent the results obtained by simulations, the squares in Fig. \ref{fig:Ex1_2} are found by using the approximation for intra-cell interference given in Lemma 5, and the squares in Fig. \ref{fig:Ex1_3} by using the bounds for the LTs given in Lemma 4. The close match between the analysis, approximations, and simulation results validates the developed mathematical model and verifies the accuracy of the assumptions in Section III-A, as well as the bound presented in Lemmas 4-5.

Several insights can be obtained from Fig .4. For instance, the figure shows that the CCUs have better performance compared to the CEUs in all cases, which is intuitive due to the larger service distances that lead to higher path-loss attenuation for CEUs compared to the CCUs. Note that the CEUs do not have sufficient power to invert their path-loss in the UL direction, and hence, the received power at the serving BS is less than $\rho$, which leads to the deteriorated UL CEU performance when compared to the CCU case. The figure also shows that there exist an optimal value of partial overlap $0 <\alpha <1$ that maximizes the UL transmission rate\footnote{The UL performance is maximized at $\alpha=0.28859$ due to the orthogonality between the used pulse shapes at this particular value, for more details refer to \cite{AlAmmouri2015Inband}.}. Hence, despite the efficient SIC (-110 dB), neither HD nor FD are optimal in the UL case due to the prominent DL interference. On the other hand, the DL performance is mainly affected by the SIC rather than the UL interference\footnote{In the case of perfect SI or very low values of $\beta_{\rm d}$, e.g. $\beta_{\rm d}=-110$ dBm, the degradation in the SINR is only due to the UL-to-DL interference, and since this is negligible compared to the DL-to-DL interference for a realistic set of network parameters, the increase in BW (linearly) overcome the decrease in the SINR which results in approximately linear curve.}. Particularly, for UE with efficient SIC, the full overlap (i.e., FD) is the best strategy for the DL. On the other hand, partial overlap is better for UE with inefficient SIC. It is worth mentioning that the SI has more prominent effect on the CEU than CCU, in which the SI nearly nullifies the DL rate for high value of $\alpha$ and efficient SIC. This is because CEUs transmit with their maximum power, which makes the residual SI power more prominent compared to CCUs.

Comparing Fig. \ref{fig:Ex1}.a and Fig. \ref{fig:Ex1}.b, we can see that the 3NT achieves close performance to the 2NT with sufficient SIC, and outperforms the 2NT with poor SIC. Note that 3NT UEs operates in HD mode and hence are not affected by SIC as shown in Fig. \ref{fig:Ex1_2}.
\begin{figure*}[t!]
    \centering
    \begin{subfigure}[b]{0.32\textwidth}
        \centering
        \includegraphics[width=2.2in]{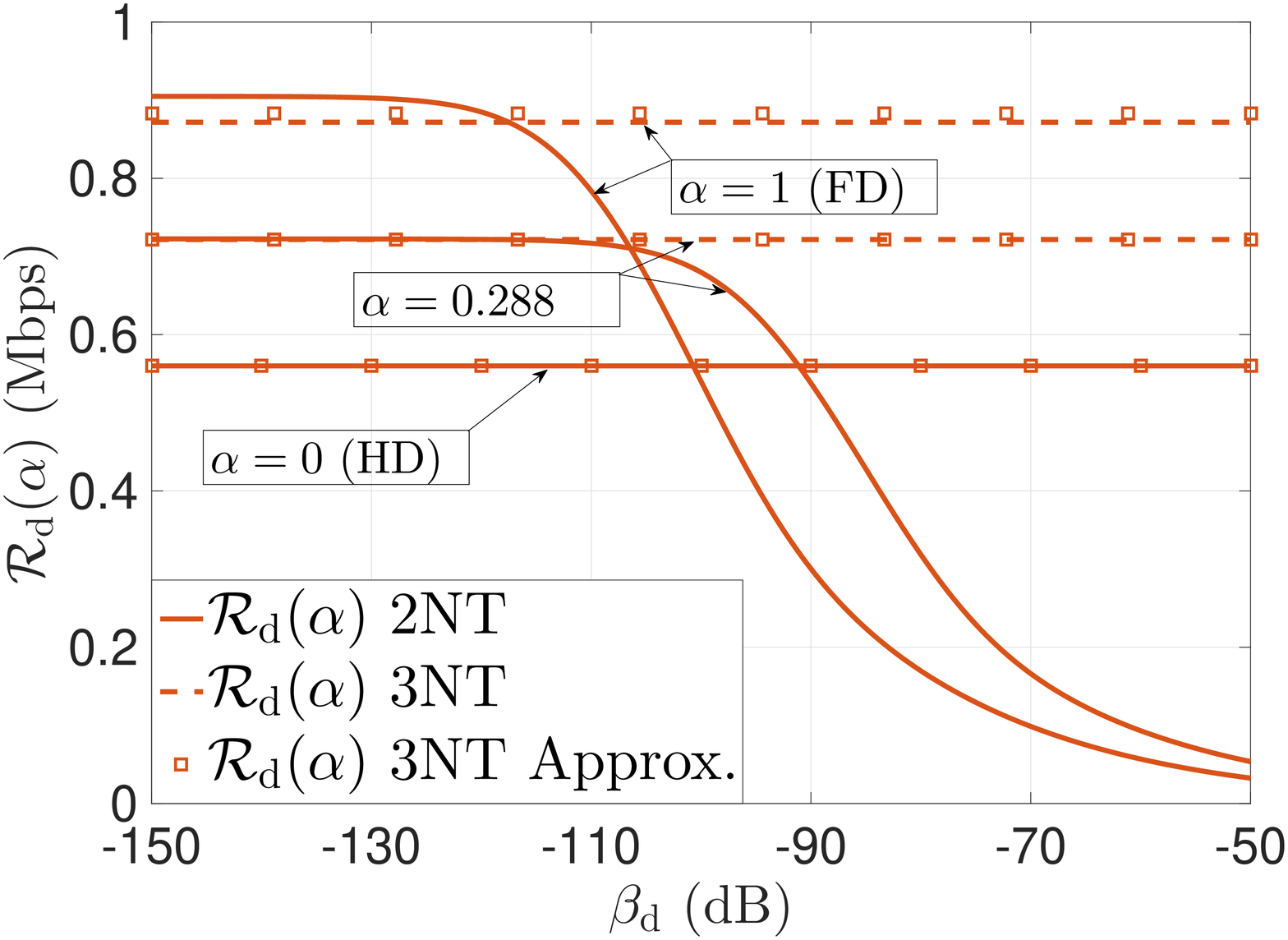}
        \caption{Average DL rate.}\label{fig:Ex2_1}
    \end{subfigure}%
    ~ 
    \begin{subfigure}[b]{0.32\textwidth}
        \centering
        \includegraphics[width=2.2in]{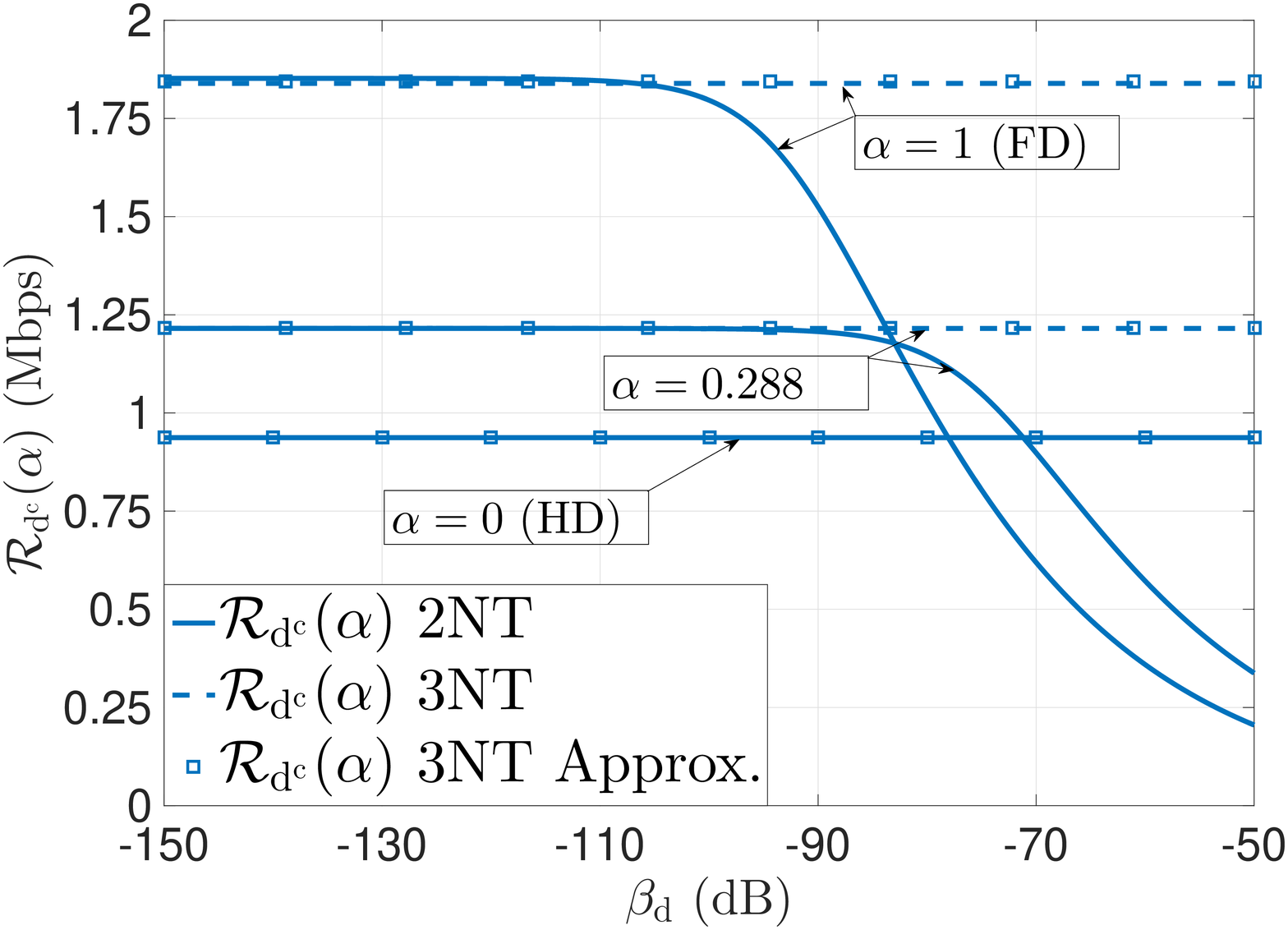}
        \caption{CCU DL  rate.}\label{fig:Ex2_2}
    \end{subfigure}
    ~ 
    \begin{subfigure}[b]{0.32\textwidth}
        \centering
        \includegraphics[width=2.2in]{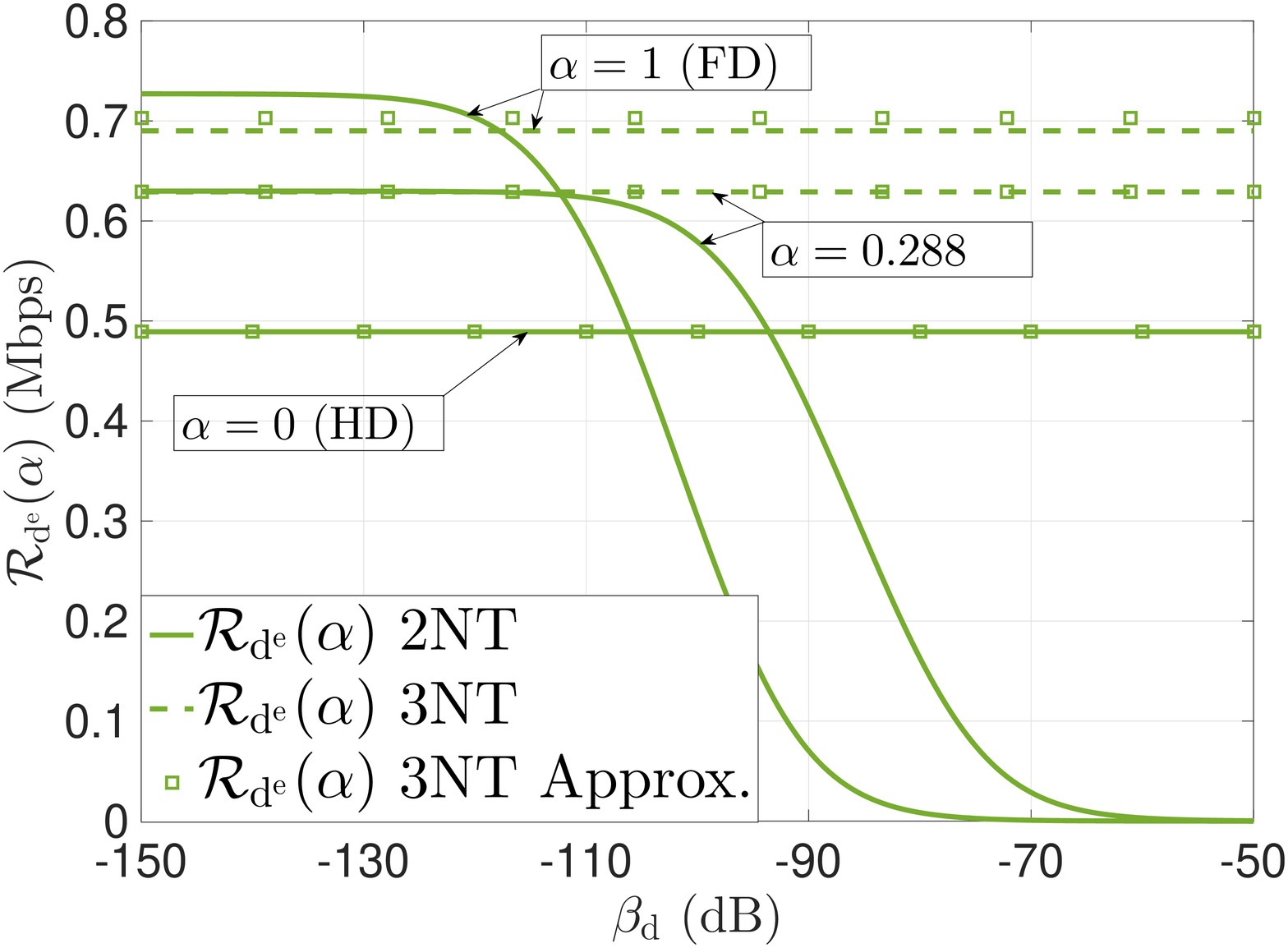}
        \caption{CEU DL rate.}\label{fig:Ex2_3}
    \end{subfigure}
    \caption{DL Rates vs $\beta_{\rm d}$ under different network topology, where FD denotes $\alpha=1$ and $\alpha$D denotes $\alpha \approx 0.2886$.}\label{fig:Ex2}\vspace{-0.8cm}
\end{figure*}

Fig.~\ref{fig:Ex2} plots the DL transmission rate vs SI attenuation power for the 2NT and 3NT. The figure shows that the DL FD rate outperforms the DL rate of both the HD and the $\alpha=0.288$ in all the cases. Fig.~\ref{fig:Ex2} also shows that there is a critical value, for $\beta_{\rm d}$, at which the 3NT outperforms 2NT. This critical value can be interpreted as the point at which the SI experienced by DL UEs in the 2NT becomes more significant than the intra-cell interference experienced by DL-UEs in 3NT and we found an closed form approximation for it in Corollary 3 and (56). Interestingly, the gain offered by the 2NT at low values of $\beta_{\rm d}$ is not significant when compared to the 3NT. Hence, the intra-cell interference is not a limiting parameter for the 3NT. In other words, network operators can harvest FD gains by HD UEs almost similar to the gains harvested by FD UEs with efficient SIC capabilities. The figure also shows that, in case of poor SIC at the UEs, the 3NT can offer significant gains, specially for CEUs, when compared to the 2NT case.

\begin{figure*}[t!]
    \centering
    \begin{subfigure}[b]{0.32\textwidth}
        \centering
        \includegraphics[width=2.2in]{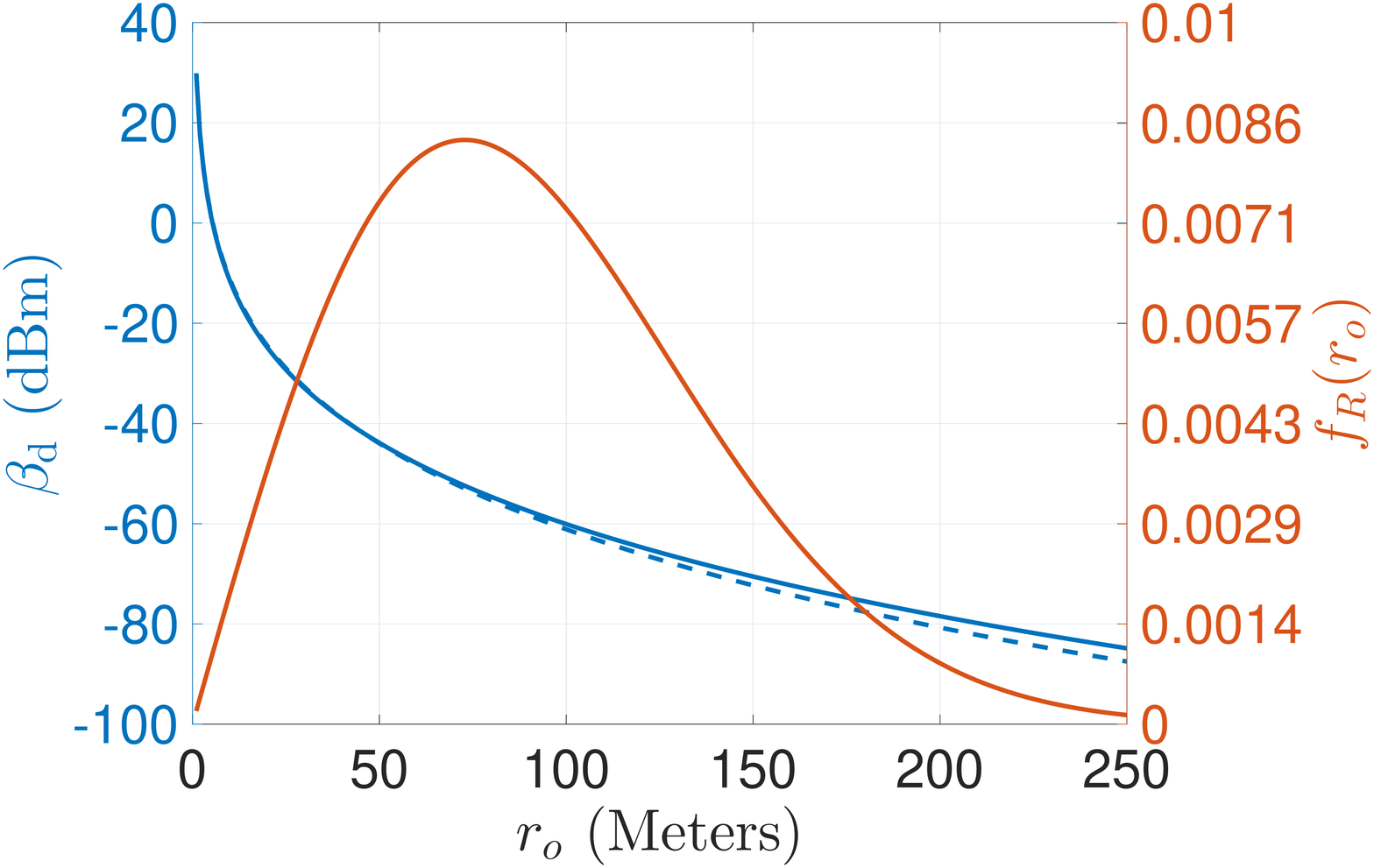}
        \caption{Critical values of $\beta_{\rm d}$  vs. the serving distance.}\label{fig:Ex3_1}
    \end{subfigure}%
    ~ 
    \begin{subfigure}[b]{0.32\textwidth}
        \centering
        \includegraphics[width=2.2in]{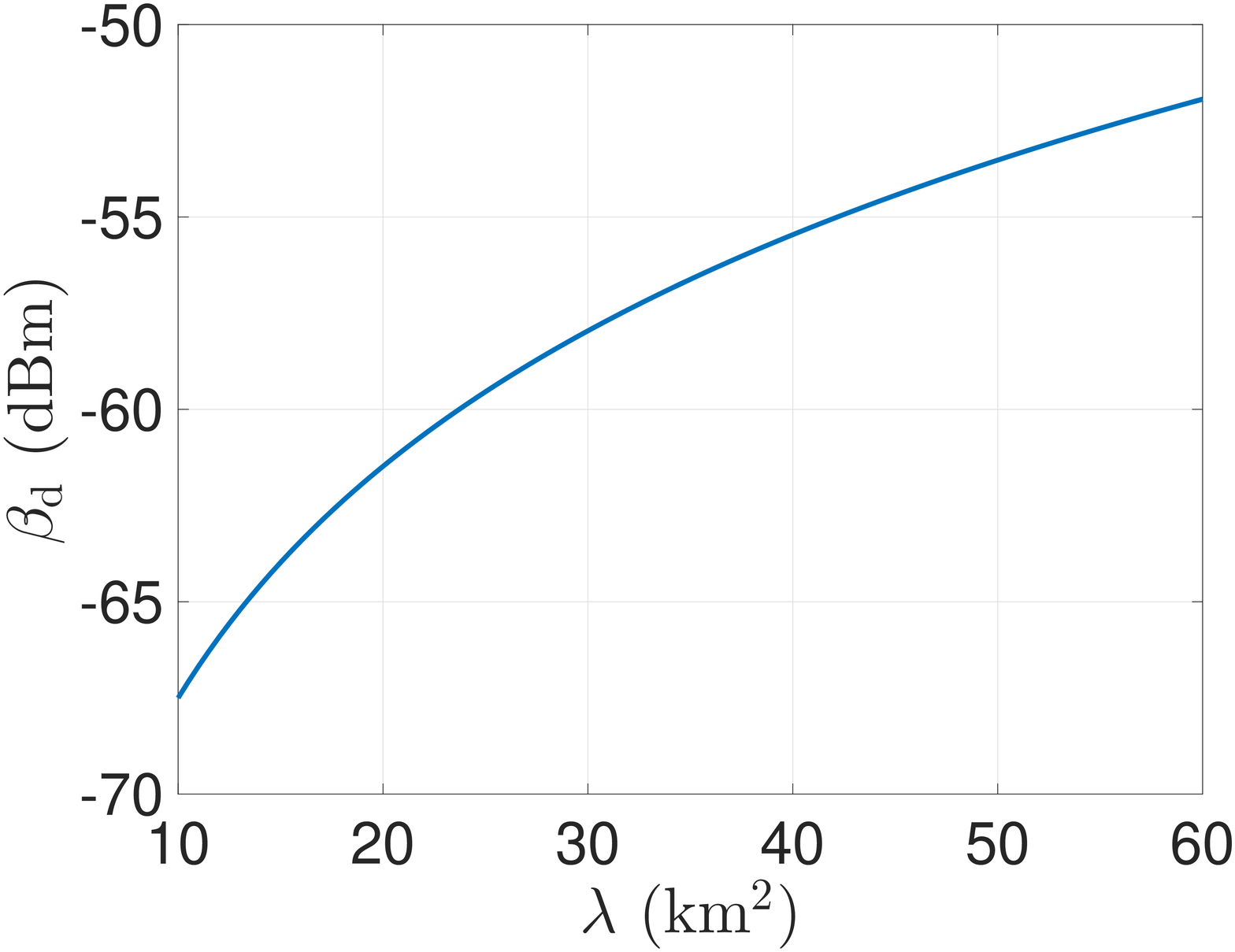}
        \caption{Critical values of $\beta_{\rm d}$  vs. the BSs' intesnity.}\label{fig:Ex3_2}
    \end{subfigure}
    ~ 
    \begin{subfigure}[b]{0.32\textwidth}
        \centering
        \includegraphics[width=2.2in]{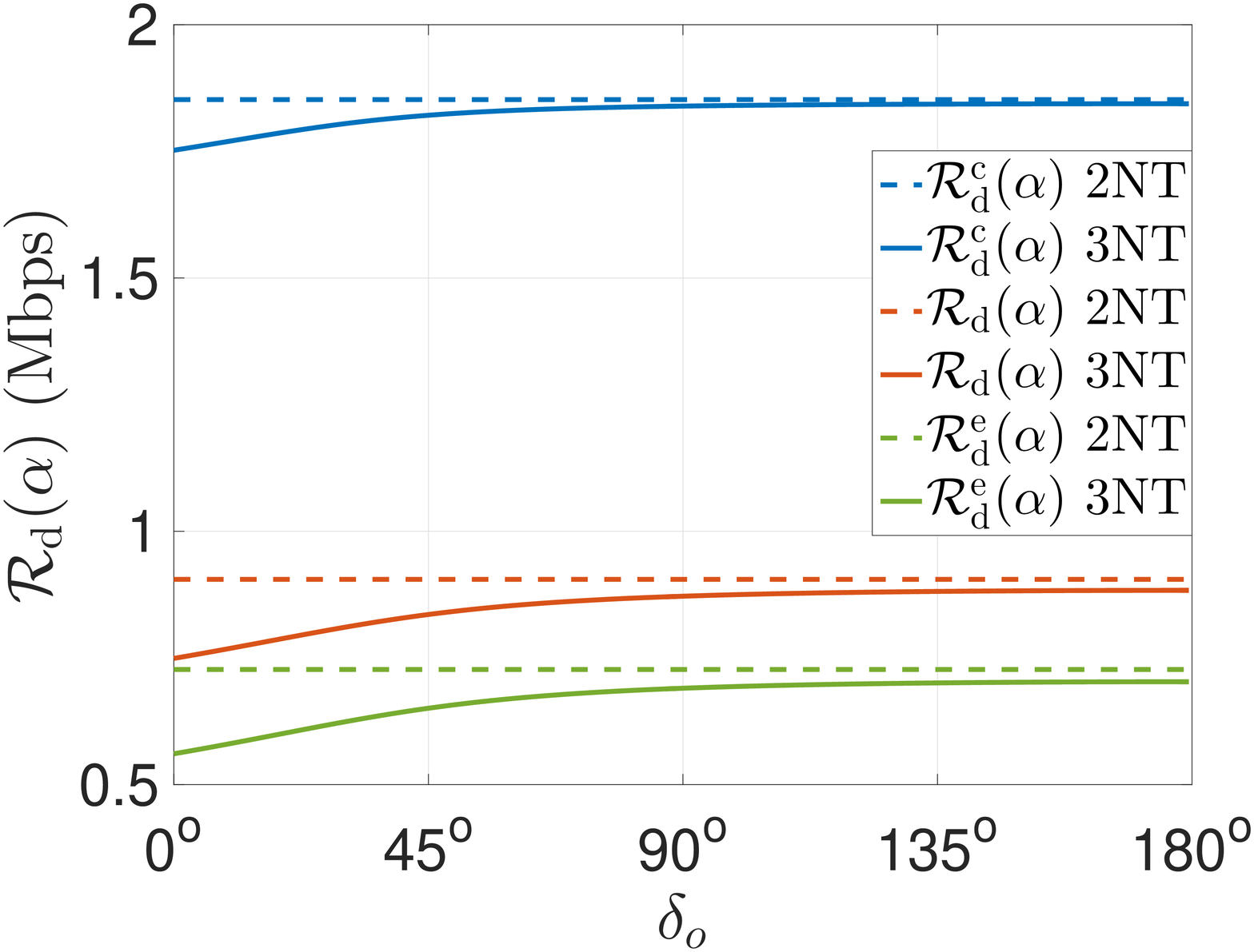}
        \caption{DL rate vs $\delta_o$ as illustrated in Fig. 2b.}\label{fig:Ex3_3}
    \end{subfigure}
    \caption{(a), (b) Critical values of $\beta_{\rm d}$  vs. the serving distance and BSs' intesnity using equations (56) and (57), and (c) DL rate vs $\delta_o$ as illustrated in Fig. 2b.}\label{fig:Ex3}\vspace{-0.8cm}
\end{figure*}

To study the effect of the serving distance on the 2NT/3NT performance, we plot Fig. \ref{fig:Ex3_1} for $\lambda=20$ BSs/km$^2$. The figure plots the minimum $\beta_{\rm d}$ required in 2NT to outperform 3NT vs. the serving distance along with the pdf of the serving distance, where sold (dashed) lines are obtained from the exact (approximate) expression of the intra-cell interference given in equation \eqref{eq:RUNT}. The close match between the exact and the approximate results validates the approximation given in \eqref{eq:RUNT}. As the figure shows, the 3NT is more appealing for farther UEs because they have a tighter constraint  for the SIC $\beta$ required for the 2NT to outperform the 3NT, which may require more sophisticated and expensive FD transceivers. There are two reasons for this result; first, large serving distance implies that the intra-cell interferer in 3NT is further on average due to the enforced scheduling technique, which reduces the negative effect of the intra-cell interference. Second, longer service distance implies larger transmit power due to the employed power control, hence more powerful SIC is required. A useful design insight for Fig. 6 is that the BSs should select the mode of operation (i.e., 3NT or 2NT) for the UEs based on their distances along with their SIC. To get more insights on the network operation for different intensities we plot Fig. \ref{fig:Ex3_2} based on equation \eqref{eq:Cond2}. As expected, 2NT becomes more appealing in dense cellular networks because the intra-cell (self) interference is more(less) prominent in smaller cell area\footnote{Fig. 6b focuses on the comparing 2NT/3NT with different intensities. The effect of intensity on the FD gain for 2NT in cellular network is covered in \cite{AlAmmouri2015Inband} and for ad-hoc network with CSMA-based transmitters is covered in \cite{Exploring2015Wang}.}.

Finally, we study the rate gains of the 2NT/3NT with $\delta_o$ in Fig. \ref{fig:Ex3_3}. As expected, by increasing $\delta_o$ the distance between the two scheduled UEs increases, hence the intra-cell power in 3NT operation decreases. Moreover, the figure shows the necessity of UEs scheduling and multi-user diversity in 3NT, otherwise the rate loss in 3NT compared to 2NT can go up to 20$\%$ in the case of random scheduling ($\delta_o=0$).

\section{Conclusion}
This paper presents a mathematical paradigm for multi-tier cellular networks with FD BSs and HD/FD UEs. The presented model captures detailed system parameters including pulse shaping, filtering, imperfect self-interference cancellation, partial uplink/downlink overlap, uplink power control, limited users' transmit powers, UE-BS association, and UEs' scheduling.  To this end, unified rate expressions for 2-nodes topology (2NT) with FD users and 3-nodes topology (3NT) with HD users are presented and used to compare their performance. The results show that there exist a critical value for the self-interference cancellation, at which the performance of 3NT outperforms the 2NT. Moreover, closed form approximations for this critical value as a function of the serving distance and the BSs' intensity are obtained. The results also show that even when SI is efficiently canceled, the 2NT does not offer significant gains when compared to the 3NT operation if multi-user diveristy and users' scheduling are exploited. This implies that network operators can harvest FD gains by implementing FD transceivers at their BSs regardless of the state of the users (i.e., FD or HD).

\appendices
\section{Proof of Lemma 1}
%Let $R_k$ denote the distance between the origin and the nearest BS in the $k^{th}$ tier,
Exploiting independence between the network tiers and using the null probability of the PPP, the cumulative distribution function (CDF) of the $i^{th}$-tier service distance is given by,
 
\small
\begin{align}\label{eq:App1_0}
&F_{R^{(i)}}(r)=\mathbb{P} \{R^{(i)} \leq r \}=\mathbb{P} \{R_i \leq r | R_i  \leq \frac{\tau_j}{\tau_i}r_j \forall j \in \{1,...,K \},j \neq i\}=\frac{\mathbb{P} \{R_i \leq r \cap R_i  \leq \frac{\tau_j}{\tau_i}r_j \forall j \neq i\}}{\mathbb{P} \{ R_i  \leq \frac{\tau_j}{\tau_i}r_j \forall j \neq i\}}
\end{align}\normalsize

The denominator is given by,
\small
\begin{align}\label{eq:App1_1}
&\mathbb{P} \{ R_i  \leq \frac{\tau_j}{\tau_i}r_j \forall j \neq i\}=\int\limits_{0}^{\infty}f_{R_i} (r_i)\left[ \prod\limits_{\underset{j=1}{j \neq i}}^{M} \ \int\limits_{(r_i \frac{\tau_i}{\tau_j})}^{\infty}f_{R_j} (r_j) dr_j \ \right]dr_i=\frac{\lambda_i}{ \sum\limits_{j=1}^{M} \frac{\tau_i^2}{\tau_j^2} \lambda_j } 
\end{align}\normalsize
and the nominator is given by,
\small
\begin{align}\label{eq:App1_2}
&\mathbb{P} \{r_i \leq r \cap r_i  \leq \frac{\tau_j}{\tau_i}r_j\forall j \neq i \}=\int\limits_{0}^{\infty}f_{R_i} (r_i)\left[ \prod\limits_{\underset{j=1}{j \neq i}}^{M}\int\limits_{r_i \frac{\tau_i}{\tau_j}}^{\infty}f_{R_j} (r_j) dr_j \ \right]dr_i= \frac{\lambda_i \left(1- \exp \left(-\pi r^2  \sum\limits_{j=1}^{M} \frac{\tau_i^2}{\tau_j^2} \lambda_j \right) \right)}{ \sum\limits_{j=1}^{M} \frac{\tau_i^2}{\tau_j^2} \lambda_j}
\end{align}\normalsize

By substituting equations \eqref{eq:App1_1} and \eqref{eq:App1_2} in \eqref{eq:App1_0} it results in,

\small
\begin{align}
F_{R^{(i)}}(r)= 1- \exp \left(-\pi r^2  \sum\limits_{j=1}^{M} \frac{\tau_i^2}{\tau_j^2} \lambda_j\right)=1- \exp \left(-\pi \bar{\lambda_i} r^2\right) \ \ \ \ \ \ \ \ \ \ \ r \geq 0.
\end{align}
where $\bar{\lambda_i}= \sum\limits_{j=1}^{M} \frac{\tau_i^2}{\tau_j^2} \lambda_j $ and the PDF is given by,
\begin{align}\label{eq:App1_3}
&f_{R^{(i)}}(r)= 2 \pi \bar{\lambda_i} r \exp \left(-\pi \bar{\lambda_i} r^2\right) \ \ \ \ \ \ \ \ \ \ \ r \geq 0.
\end{align}\normalsize

Given that the UE is a CCU, the PDF in \eqref{eq:App1_3} should be truncated according to channel inversion power control. Let $R^{(i)}_c$ denote the serving distance for a test CCU connected to the $i^{th}$ tier, then its PDF is given by,

\small
\begin{align}\label{eq:App1_4}
\!\!\!\!\!\!\!\!\!\!\!\!\!\! f_{R^{(i)}_c}(r)&= \frac{2 \pi \bar{\lambda_i} r \exp \left(-\pi \bar{\lambda_i} r^2\right)}{\int\limits_{0}^{\left(\frac{P_{\rm u}}{\rho^{(i)}}\right)^{\frac{1}{\eta_{\rm dd}}}} 2 \pi \bar{\lambda_i} r \exp \left(-\pi \bar{\lambda_i} r^2\right)dr}=\frac{2 \pi \bar{\lambda_i} r \exp \left(-\pi \bar{\lambda_i} r^2\right)}{1-\exp \left(-\pi \bar{\lambda_i} \left(\frac{P_{\rm u}}{\rho^{(i)}}\right)^{\frac{2}{\eta_{\rm dd}}}\right)} \mathbbm{1}_{\left\{0\leq r \leq \left(\frac{P_{\rm u}}{\rho^{(i)}}\right)^{\frac{1}{\eta_{\rm dd}}}\right\}}(r).
\end{align}\normalsize

Similarly, for the PDF of the service distance for a CEUs,
\small
\begin{align}\label{eq:App1_5}
\!\!\!\!\!\!\!\!\!\!\!\!\!\! f_{R^{(i)}_e}(r)&=\frac{2 \pi \bar{\lambda_i} r \exp \left(-\pi \bar{\lambda_i} r^2\right)}{\int\limits_{\left(\frac{P_{\rm u}}{\rho^{(i)}}\right)^{\frac{1}{\eta_{\rm dd}}}}^{\infty} 2 \pi \bar{\lambda_i} r \exp \left(-\pi \bar{\lambda_i} r^2\right)dr}= 2 \pi \bar{\lambda_i} r \exp \left(-\pi \bar{\lambda_i} r^2+\pi \bar{\lambda_i} \left(\frac{P_{\rm u}}{\rho^{(i)}}\right)^{\frac{2}{\eta_{\rm dd}}}\right) \mathbbm{1}_{\left\{\left(\frac{P_{\rm u}}{\rho^{(i)}}\right)^{\frac{1}{\eta_{\rm dd}}} < r < \infty\right\}}(r).
\end{align}\normalsize

\section{Proof of Lemma 2}
The proof is as follows,
\small
\begin{align}
   \mathcal{L}_{I}(s) &= \mathbb{E}\left[\exp \left( \sum\limits_{j \in \Phi}-s P_{_j} h_j r_j^{-\eta} \mathbbm{1}\left( r_j>a_j\right) \right)\right],\notag \\
&\stackrel{(i)}{=}  \mathbb{E}_{\Phi}\left[\underset{r_j \in \Phi}{\prod} \mathbb{E}_{h_j,P_{j}}\left[ \exp\left( -s P_{_j} h_j r_j^{-\eta} \mathbbm{1}\left( r_j> a_j\right) \right)\right]\right],\notag \\
 &\stackrel{(ii)}{=}  \exp\left( - 2 \pi   \lambda\mathbb{E}_{P}\left[ \int_{a}^{\infty}\mathbb{E}_{h}\left[  \left(1- \exp\left( -s P h r^{-\eta}  \right)\right)\right] rdr\right] \right),\notag \\
 &\stackrel{(iii)}{=} \exp \left( \frac{-2 \pi \lambda}{\eta-2} \mathbb{E}_{P} \left[a^{2-\eta} s P \  {}_2 F_1 \left[1,1-\frac{2}{\eta},2-\frac{2}{\eta}, -a^{-\eta} P s \right]  \right] \right),\label{eq:LUU}
\end{align}\normalsize
where, $(i)$ follows from the independence between $\tilde{\Psi}$ and $h_j$, $(ii)$ by using the probability generation functional (PGFL) of PPP and $(iii)$ by using the LT of $h$ and by evaluating the integral.

\section{Proof of Lemma 3}
The lemma is proved by showing that the second derivative of the function which appears inside the expectation of the exponent in \eqref{eq:LTgeneralapp} is positive  w.r.t P. Hence, the function of interest is  convex in P and the result in Lemma 3 follows from Jensen's inequality \cite[Section 3.1.8]{Convex2004Boyd}. Let $y=a^{-\eta}P s$,  the function of interest, denoted here as $G(y)$, can be expressed as
\small
\begin{align} \label{gyyg}
G(y)=-y \  {}_2 F_1 \left[1,1-\frac{2}{\eta},2-\frac{2}{\eta}, -y \right]
\end{align}
\normalsize
The second derivative of $G(y)$ is given by
\footnotesize
\begin{align}\label{2nd_der}
\frac{d^2 G(y)}{dy^2}&%\stackrel{(i)} {=}-\left(1-\frac{2}{\eta }\right) \left(-\frac{\left(1-\frac{2}{\eta }\right) \left(\frac{1}{y+1}-\, _2F_1\left(1,1-\frac{2}{\eta };2-\frac{2}{\eta };-y\right)\right)}{y}-\frac{1}{(y+1)^2}\right) \notag \\
%& \ \ \ \ \ \ \ \ \ \ -\frac{\left(1-\frac{2}{\eta }\right) \left(\frac{1}{y+1}-\, _2F_1\left(1,1-\frac{2}{\eta };2-\frac{2}{\eta };-y\right)\right)}{y}, \notag \\
{=}\left(\frac{\left(\, _2F_1\left(1,1-\frac{2}{\eta };2-\frac{2}{\eta };-y\right)-\frac{1}{y+1}\right)}{y}\left(-\frac{2}{\eta }\right)+\frac{1}{(y+1)^2} \right)\left(1-\frac{2}{\eta }\right).
\end{align}\normalsize
where \eqref{2nd_der} is found by using \cite[Eqs (15.2.2),(15.2.10),(15.2.27)]{Handbook1964Abramowitz} and some mathematical simplifications. Owing to the fact that $\frac{1}{(1+y)^2}$, $\left(1-\frac{2}{\eta }\right)$, $\frac{2}{\eta}$, and y are positive for $\eta>2$, the prove is completed by proving that  $G_2(y)= \left(\, _2F_1\left(1,1-\frac{2}{\eta };2-\frac{2}{\eta };-y\right)-\frac{1}{y+1}\right)$ is positive. Using the integral definition of the hypergeometric function \cite[Eq. (15.3.1)]{Handbook1964Abramowitz} and projecting it on our case, we have
\small
\begin{align}\label{eq:App3_1}
 \!\!\!\!\!\!\!\!\!\!\!\!\!{}_2F_1\left(1,1-\frac{2}{\eta };2-\frac{2}{\eta };-y\right)&= \frac{\Gamma \left(2-\frac{2}{\eta} \right)}{\Gamma \left(1-\frac{2}{\eta} \right)} \int\limits_0^{1} \frac{t^{\frac{-2}{\eta}}}{1+t y}dt \stackrel{(i)}{=} \left(1-\frac{2}{\eta} \right) \int\limits_0^{1} \frac{t^{\frac{-2}{\eta}}}{1+t y}dt\stackrel{(ii)}{>} \int\limits_0^{1} \frac{t^{\frac{-2}{\eta}}}{1+t y}dt \stackrel{(iii)}{>} \frac{1}{1+y} > 0.
\end{align} \normalsize
where $(i)$ follows by \cite[Eq. (6.1.15)]{Handbook1964Abramowitz}, $(ii)$ follows from the fact that $\eta>2$, and (iii)  is proved as in the sequel.  Taking the first derivative of the integrand in \eqref{eq:App3_1} as
\small
\begin{align}
{\left(\frac{t^{-2/\eta }}{t y+1}\right)}^{\prime}=\frac{t^{-\frac{\eta +2}{\eta }} (-(\eta +2) t y-2)}{\eta  (t y+1)^2},
\end{align} \normalsize
shows that it is a decreasing function in $t$, and hence, the minimum occurs at the boundary $1$. Then (iii) in \eqref{eq:App3_1} follows by lower-bounding the integral by the minimum value of the integrand multiplied by the integration region. Hence, the second derivative of $G(y)$ in \eqref{gyyg} is positive, which completes the prove.
%\small
%\begin{align}
 %\int\limits_0^{1} \frac{t^{\frac{-2}{\eta}}}{1+t y}dt> \frac{1}{1+y}.
%\end{align} \normalsize

%By substituting the inequality in $G_2(y)$ we get,
%\small
%\begin{align}
%G_2(y)>\frac{1}{1+y}-\frac{1}{1+y}=0.
%\end{align}\normalsize
%hence, the second derivative is positive everywhere is terms of $y$, or equivalently the exponent in \eqref{eq:LTgeneralapp} is convex w.r.t $P$ and Jensen's inequality can be applied which results in,
%\small
%\begin{align}
%\!\!\!\!\!\!\!\!\!\!\!\! \frac{-2 \pi \lambda}{\eta-2} a^{2-\eta} s \mathbb{E} \left[P\right] \  {}_2 F_1 \left[1,1-\frac{2}{\eta},2-\frac{2}{\eta}, -a^{-\eta} \mathbb{E} \left[P\right] s \right]  \leq \frac{-2 \pi \lambda}{\eta-2} \mathbb{E}_{P} \left[a^{2-\eta} s P \  {}_2 F_1 \left[1,1-\frac{2}{\eta},2-\frac{2}{\eta}, -a^{-\eta} P s \right]  \right],
%\end{align}\normalsize

\section{Proof of Lemma 4}

Based on Lemma 2 and 3, we only need to determine the interference exclusion region (IER) for each tier  ($a_j$) in each case. 
\begin{itemize}
\item $\mathcal{L}_{I^{(k,i)}_{{\rm d}\rightarrow {\rm d}}}(s)$: Due to the association rule in Section~\ref{sec:Association}, $r_o \tau_i \leq r_j \tau_k$ is always satisfied. Hence, the IER is defined by $\mathcal{B}(o,r_o \frac{\tau_i}{\tau_k})$, by substituting $a$ in Lemma 2 by $r_o \frac{\tau_i}{\tau_k}$, the final expression is found.

\item $\mathcal{L}^{(c)}_{I^{(k,i)}_{{\rm d}\rightarrow {\rm d}}}(s)$: Based on the power inversion for CCUs and following \cite{On2014ElSawy}, $a=(\frac{P_{\rm u}}{\rho^{(k)}})^{\frac{1}{\eta}}$.

\item $\mathcal{L}^{(e)}_{I^{(k,i)}_{{\rm d}\rightarrow {\rm d}}}(s)$: Based on the power inversion for CEUs and following \cite{Load2014AlAmmouri}, $a=r_o$, then by using Lemma 3, the final expression is found.

\item $\mathcal{L}_{I^{(k,i)}_{{\rm d}\rightarrow {\rm u}}}(s)$: The PPP assumption of the BSs location implies that there is no IER for both cases (CCUs and CCUs), and hence $a=0$.

\item $\mathcal{L}_{I^{(k,i)}_{{\rm u}\rightarrow {\rm d}}}(s)$: We assume that the tagged BS is collocated its associated UE, hence $\mathcal{L}_{I^{(k,i)}_{{\rm u}\rightarrow {\rm d}}}(s)=\mathcal{L}_{I^{(k,i)}_{{\rm u}\rightarrow {\rm u}}}(s)$ in 2NT. In 3NT, the effect of intra-cell interference should be considered also. Let $U^{(k,i)}_1(s)$ denote the LT of the intra-cell interference and let $P_{{\rm u}_1}$, $h_{1-o}$, $r_{1-o}$, and $r_{1}$ denote the transmitted power of the interfering user, the channel gain between the two users, the distance between them and the distance between the interfering UE and the serving BS, respectively, then the LT of the interfering power can be expressed as
\small
\begin{align}\label{eq:E}
 U^{(i,i)}_1(s)&=\mathbb{E}\left[ e^{-s P_{{\rm u}_1} h_{1-o} r_{1-o}^{-\eta} }\right], \notag \\
 &\stackrel{(i)}{=}\mathbb{E}\left[ e^{-s P_{{\rm u}_1} h_{1-o} \left(r_o^2+r_1^2-2 r_o r_1 \cos (\delta)\right)^{-\eta/2} }\right],
\end{align}\normalsize
where $(i)$ follows by using the cosine rule (cf. Fig. 2b), where $\delta$ is the uniformly distributed between $\delta_o$ and $\pi$. When the other UE is a CCU, which has a probability $\mathbb{P}\{\rm CCU \}$, then $P_{{\rm u}_1}=\rho^{(i)} r_1^{\eta}$, and when it is CEU, which has a probability $\mathbb{P}\{\rm CEU \}$, then $P_{{\rm u}_1}=P_{\rm u}$ by substituting these values and by averaging over $h_{1-o}$, the expression for $ U^{(k,i)}_1(s)$ is found.
\end{itemize}

\section{Proof of Theorem 1}
Starting by the UL outage probability, for CCUs the transmitted power is equal to $\rho r_o^{\eta}$, and for the CEUs the transmitted power is set to the maximum $P_{\rm u}$, by substituting these values in \eqref{eq:outageGeneral3} we get equations \eqref{eq:OutageUL1} and \eqref{eq:OutageUL2} except $ U^{(i)}_{\rm SI_ u}$ which is found by substituting $\tilde{\sigma}^2_{s}$ by its value given in \eqref{eq:SIu2} and then by averaging over $h_s$ while conditioning on $r_o$. Similar steps are followed to find the outage in the DL direction.

\bibliographystyle{IEEEtran}
\bibliography{ref}

\vfill

\end{document}